\documentclass[sigconf, natbib=true, dvipsnames]{acmart}

\usepackage{kotex}
\usepackage{dsfont}
\usepackage{bbm}
\usepackage{array}
\usepackage{arydshln}
\usepackage{enumitem}
\usepackage{hyperref}
\usepackage{setspace}
\hypersetup{
    colorlinks=true,
    linkcolor=blue,
    filecolor=magenta,      
    urlcolor=cyan,
    }

\usepackage{booktabs,multirow}
\usepackage{algorithmic}
\usepackage[linesnumbered,algoruled,boxed,lined]{algorithm2e}
\usepackage{balance}

\newcolumntype{P}[1]{>{\centering\arraybackslash}p{#1}}

\newcommand{\ie}{{\it i.e.}}
\newcommand{\eg}{{\it e.g.}}
\newcommand{\topN}{{top-\emph{N}}}
\newcommand{\com}{\textcolor{red}}
\newcommand{\rev}{\textcolor{blue}}
\newcommand{\green}{\textcolor{Green}}

\newtheorem*{assumption*}{\assumptionnumber}
\providecommand{\assumptionnumber}{}


\AtBeginDocument{%
  \providecommand\BibTeX{{%
    \normalfont B\kern-0.5em{\scshape i\kern-0.25em b}\kern-0.8em\TeX}}}

\copyrightyear{2023}
\acmYear{2023}
\setcopyright{acmlicensed}
\acmConference[CIKM '23]{Proceedings of the 32nd ACM International Conference on Information and Knowledge Management}{October 21--25, 2023}{Birmingham, United Kingdom}
\acmBooktitle{Proceedings of the 32nd ACM International Conference on
Information and Knowledge Management (CIKM '23), October 21--25, 2023,
Birmingham, United Kingdom}
\acmPrice{15.00}
\acmDOI{10.1145/3583780.3615086}
\acmISBN{979-8-4007-0124-5/23/10}

\begin{document}

\title{Toward a Better Understanding of Loss Functions for Collaborative Filtering}

\author{Seongmin Park}
\affiliation{
  \institution{Sungkyunkwan University\\Suwon, Republic of Korea}
  }
\email{psm1206@skku.edu}

\author{Mincheol Yoon}
\affiliation{
  \institution{Sungkyunkwan University\\Suwon, Republic of Korea}
  }
\email{yoon56@skku.edu}

\author{Jae-woong Lee}
\affiliation{
  \institution{Sungkyunkwan University\\Suwon, Republic of Korea}
  }
\email{jwlee.icc@skku.edu}

\author{Hogun Park}
\affiliation{
  \institution{Sungkyunkwan University\\Suwon, Republic of Korea}
  }
\email{hogunpark@skku.edu}

\author{Jongwuk Lee}\authornote{Corresponding author}
\affiliation{
  \institution{Sungkyunkwan University\\Suwon, Republic of Korea}
  }
\email{jongwuklee@skku.edu}

\renewcommand{\shortauthors}{Seongmin Park, Mincheol Yoon, Jae-woong Lee, Hogun Park, \& Jongwuk Lee}

\begin{abstract}

Collaborative filtering (CF) is a pivotal technique in modern recommender systems. The learning process of CF models typically consists of three components: \emph{interaction encoder}, \emph{loss function}, and \emph{negative sampling}. Although many existing studies have proposed various CF models to design sophisticated interaction encoders, recent work shows that simply reformulating the loss functions can achieve significant performance gains. This paper delves into analyzing the relationship among existing loss functions. Our mathematical analysis reveals that the previous loss functions can be interpreted as \emph{alignment} and \emph{uniformity} functions: (i) the alignment matches user and item representations, and (ii) the uniformity disperses user and item distributions. Inspired by this analysis, we propose a novel loss function that improves the design of alignment and uniformity considering the unique patterns of datasets called \emph{Margin-aware Alignment and Weighted Uniformity (MAWU)}. The key novelty of MAWU is two-fold: (i) \emph{margin-aware alignment (MA)} mitigates user/item-specific popularity biases, and (ii) \emph{weighted uniformity (WU)} adjusts the significance between user and item uniformities to reflect the inherent characteristics of datasets. Extensive experimental results show that MF and LightGCN equipped with MAWU are comparable or superior to state-of-the-art CF models with various loss functions on three public datasets.

\end{abstract}

\ccsdesc[500]{Information systems~Recommender systems}
\ccsdesc[500]{Information systems~Collaborative filtering}

\keywords{Collaborative filtering, loss function, theoretical analysis, alignment and uniformity}

\maketitle

\section{Introduction}\label{sec:Introduction}

\begin{table*}[t]
\centering
\caption{Architecture comparison between CF losses categorized by four types. UIB~\cite{ZhuoZYZ22} uses BPR~\cite{RendleFGS09} as its backbone. All the notations in loss functions follow the notations of each original paper. Due to limited space, DirectAU is denoted as DAU. 
}
\vspace{-4mm}

\label{tab:architecture_losses}
\begin{center}
\renewcommand{\arraystretch}{1.8} 
\begin{tabular}{P{1.8cm}|P{1.15cm}|P{3.3cm}|P{10.0cm}}
\toprule
 Type & Name & Similarity or distance & Loss function \\
 \hline
 \multirow{3}{*}{Pointwise}
  & BCE~\cite{HuKV08} & $s(u,i)=\Tilde{f(u)}^\text{T}\Tilde{f(i)}$ & $-\sum_{(u, i) \in \mathcal{D}} \log \sigma \left(s(u, i)\right) - \sum_{(u, j) \notin \mathcal{D}} \log \left(1 - \sigma (s(u, j))\right)$  \\
  & MCL~\cite{GaoCPSV22} & $d(u,i)=||\Tilde{f(u)}-\Tilde{f(i)}||_2^2$ & $ \frac{1}{\alpha} \log \left(1+\frac{1}{m} \sum_{(u, i) \in \mathcal{D}} e^{\alpha\left(d(u, i)+\lambda_p\right)}\right) + \frac{1}{\beta} \log \left(1+\frac{1}{m} \sum_{(u, j) \notin \mathcal{D}} e^{-\beta\left(d(u, j)+\lambda_n\right)}\right) $  \\
  & UIB~\cite{ZhuoZYZ22} & $s(u,i)=\Tilde{f(u)}^\text{T}\Tilde{f(i)}$ & $-\sum_{(u, i) \in \mathcal{D}} \log \sigma\left(s(u, i)-b_u\right)-\alpha \sum_{(u, j) \notin \mathcal{D}} \log \sigma\left(b_u-s(u, j)\right)$  \\
 \hline
 \multirow{3}{*}{Pairwise}
  & BPR~\cite{RendleFGS09} & $s(u,i)=\Tilde{f(u)}^\text{T}\Tilde{f(i)}$ &  $ - \sum_{(u, i, j) \in \mathcal{T}} \log \sigma\left( s(u,i)-s(u,j) \right) $ \\
  & CML~\cite{HsiehYCLBE17} & $d(u,i)=||\Tilde{f(u)}-\Tilde{f(i)}||_2^2$ & $ \sum_{(u, i, j) \in \mathcal{T}} \left[ d(u,i)-d(u,j) + M \right]_+ $  \\
  & SML~\cite{LiZZQZHH20} & $d(u,i)=||\Tilde{f(u)}-\Tilde{f(i)}||_2^2$ & $ \sum_{(u, i, j) \in \mathcal{T}} \left( \left[ d(u,i)-d(u,j) + M_u \right]_+ + \left[ d(u,i)-d(i,j) + M_i \right]_+\right) + \lambda \mathcal{L}_{AM} $  \\
 \hline
 \multirow{3}{*}{Setwise}
  & CCL~\cite{MaoZWDDXH21} & $s(u,i)=\Tilde{f(u)}^\text{T}\Tilde{f(i)}$ & $ \sum_{(u,i,\mathcal{N}_u) \in \mathcal{T}_s} \left( 1-s(u,i) + \frac{w}{|\mathcal{N}_u|} \sum_{j \in \mathcal{N}_u} [s(u,j)-M]_+ \right)$  \\
  & SSM~\cite{WuWGCFQH22} & $s(u,i)=\Tilde{f(u)}^\text{T}\Tilde{f(i)}$ & $ -\sum_{(u,i,\mathcal{N}_u) \in \mathcal{T}_s} \log \frac{ \exp (s(u,i) / \tau) }{ \exp(s(u,i) / \tau) + \sum_{j \in \mathcal{N}_u} \exp (s(u,j) / \tau)} $  \\
  & BC~\cite{ZhangMWC22} & $s(u,i)=\cos(\hat{\theta}_{ui})$ & $ -\sum_{(u,i,\mathcal{N}_u) \in \mathcal{T}_s} \log \frac{\exp (\cos (\hat{\theta}_{u i}+M_{u i}) / \tau) }{\exp (\cos (\hat{\theta}_{u i}+M_{u i}) / \tau) + \sum_{j \in \mathcal{N}_u} \exp (\cos (\hat{\theta}_{u j}) / \tau)} $ \\
 \hline
 Align. \& Unif. & DAU~\cite{WangYMZCLM22} & $d(u,i)=||\Tilde{f(u)}-\Tilde{f(i)}||_2^2$ & $ \mathbb{E}_{\left(u, i\right) \sim p_{pos}} d(u,i) 
 + \gamma \left(\log \mathbb{E}_{\left(u, u^{\prime}\right) \sim p_{user}} e^{-2 \cdot d\left(u, u^{\prime}\right)}+\log \mathbb{E}_{\left(i, i^{\prime}\right) \sim p_{item}} e^{-2 \cdot d\left(i, i^{\prime}\right)} \right) $  \\
 
\bottomrule
\end{tabular}
\end{center}
\vspace{-4mm}
\end{table*}

Recommender systems~\cite{RicciRS15} are widespread in various Web applications (e.g., YouTube, Amazon, Netflix, and Spotify) to assist users in navigating the abundance of information available during the decision-making process. Collaborative filtering (CF)~\cite{RendleFGS09, HeLZNHC17, SedhainMSX15, HuKV08, PanZCLLSY08, LiangKHJ18, WuDZE16, ZhuWC19, Koren08, 0001DWLZ020, WangWY15, XueDZHC17} is a core technique for identifying meaningful collaborative signals among users/items and predicting hidden user preferences on items. The primary advantage of CF is that it only uses past user-item interactions without the need for auxiliary information about users/items. Owing to its simplicity and efficacy, numerous CF models have been developed for various domains and tasks in recommender systems.

The learning process of CF models primarily consists of three key components: \emph{interaction encoder}, \emph{loss function}, and \emph{negative sampling}~\cite{MaoZWDDXH21}. Many studies~\cite{NingK11, Steck19, Koren08, HuKV08, SalakhutdinovM07, ChenZZLM20, SedhainMSX15, LiangKHJ18, ShenbinATMN20, 0001DWLZ020, WuWF0CLX21, YuY00CN22, CaiHXR23} have focused on designing effective interaction encoders to capture complex correlations across users and items. However, some works~\cite{DacremaCJ19, ChinCC22, abs-2305-01801} have pointed out that the performance of these encoders may be overestimated under certain experimental conditions. Notably, recent studies \cite{MaoZWDDXH21, WuWGCFQH22, WangYMZCLM22, ZhangMWC22} have highlighted that reformulating loss functions is significant for improving the performance of CF models. In particular, MF~\cite{HuKV08} equipped with CCL~\cite{MaoZWDDXH21}, SSM~\cite{WuWGCFQH22}, DirectAU~\cite{WangYMZCLM22}, and BC~\cite{ZhangMWC22} loss functions has yielded competitive performance compared to state-of-the-art CF models. However, there is no previous study that represents an in-depth analysis of existing loss functions mathematically.

Before diving into the analysis, we first categorize existing loss functions~\cite{HuKV08, RendleFGS09, LiZZQZHH20, GaoCPSV22, MaoZWDDXH21, WuWGCFQH22, WangYMZCLM22, ZhuoZYZ22, ZhangMWC22, HsiehYCLBE17} into four types, \ie, \emph{pointwise}, \emph{pairwise}, \emph{setwise}, and \emph{alignment and uniformity (AU)} loss functions, as summarized in Table~\ref{tab:architecture_losses}. The pointwise loss functions are divided into two loss terms according to positive and negative labels of user-item interactions. The pairwise and setwise loss functions employ a single loss term with a triplet $\langle$user, positive~item, negative~item(s)$\rangle$. While the pairwise loss function uses a single negative item, the setwise loss function utilizes a set of negative items in one triplet. Meanwhile, the AU loss function receives positive user-item pairs to compute the alignment function and utilizes all user and item samples in a batch to compute the uniformity function.

Then, we investigate the mathematical relationships among the existing loss functions. We mainly consider seven loss functions such as BPR~\cite{RendleFGS09}, CML~\cite{HsiehYCLBE17}, SML~\cite{LiZZQZHH20}, CCL~\cite{MaoZWDDXH21}, SSM~\cite{WuWGCFQH22}, BC~\cite{ZhangMWC22}, and DirectAU~\cite{WangYMZCLM22} except for the pointwise loss functions\footnote{The pointwise loss functions are difficult to integrate mathematically with other loss functions because they are split into two loss terms based on the labels.}. Our analysis reveals the following key observations: (i) SSM~\cite{WuWGCFQH22} is regarded as the fundamental loss function among pairwise and setwise loss functions. Moreover, BC~\cite{ZhangMWC22} represents an improvement over the loss function of SSM~\cite{WuWGCFQH22} by introducing the interaction-specific angular margins. (ii) SSM~\cite{WuWGCFQH22} is also interpreted as DirectAU~\cite{WangYMZCLM22}, consisting of two loss functions, \ie, alignment and uniformity for contrastive representation learning. These observations highlight two main ways in which the performance of recent recommendation systems can be improved.
First, BC~\cite{ZhangMWC22} can learn angular margins to account for the popularity bias inherent in each user-item interaction. Second, DirectAU~\cite{WangYMZCLM22} does not require negative sampling strategies, which saves additional costs.

The above analysis has led us to design an improved loss function, called \emph{Margin-aware Alignment and Weighted Uniformity} (\emph{MAWU}), leveraging the strengths of both BC~\cite{ZhangMWC22} and DirectAU~\cite{WangYMZCLM22}. Specifically, it consists of the following two loss functions.

\begin{itemize}[leftmargin=5mm]
\item \textbf{Margin-aware alignment (MA)}: It enhances the existing alignment function by introducing user/item-specific margins. It can further mitigate the biases from the active users and popular items, requiring fewer parameters during the learning. As a result, more robust predictions are expected, particularly on sparser interaction graphs.

\item \textbf{Weighted uniformity (WU)}: It employs different hyperparameters to adjust the significance of user and item uniformities. For instance, if the user distribution is more uniform than the item distribution, we emphasize user uniformity more. Despite its simplicity, it easily reflects the inherent statistics of datasets, \ie, Gini indices of user and item, improving the performance of the original uniformity function.
\end{itemize}

The main advantage of MAWU is two-fold: (i) It can learn better user/item representations by directly optimizing alignment and uniformity. (ii) It adaptively reflects the unique characteristics of datasets by using the margin for alignment and the adjusted importance for uniformity.

The key contributions of this paper are summarized as follows:
\begin{itemize}[leftmargin=5mm]
    \item Through a mathematical analysis of seven loss functions, we derive that SSM is the fundamental loss function, which is also expressed in the form of optimizing alignment and uniformity. Moreover, we suggest the key potential of combining BC~\cite{ZhangMWC22} and DirectAU~\cite{WangYMZCLM22}. (Section~\ref{sec:analysis})

    \item We propose a \emph{Margin-aware Alignment and Weighted Uniformity (MAWU)} loss function. (i) Margin-aware alignment (MA) efficiently learns with the angular margin for mitigating user/item popularity biases, and (ii) weighted uniformity (WU) is used to adjust the significance of the user/item uniformity. (Section~\ref{sec:model})

    \item Extensive experimental results demonstrate that MAWU outperforms ten existing loss functions and eight state-of-the-art CF models on three benchmark datasets. (Sections~\ref{sec:setup}-\ref{sec:result})

\end{itemize}
\section{Background}\label{sec:background}

\subsection{Formulation of CF models}

\textbf{Notations}. Let $\mathcal{U}$ and $\mathcal{I}$ denote a set of users and a set of items, respectively. Assuming the implicit setting, we define a user-item interaction matrix $\textbf{R} \in \{0,1\}^{|\mathcal{U}| \times |\mathcal{I}|}$. If $r_{ui} = 1$, it indicates user $u$ has interacted with item $i$. Otherwise, \emph{i.e.}, $r_{ui} = 0$, there is no interaction between user $u$ and item $i$. Given user $u$, we define $\mathcal{I}_u^+ = \{i \in \mathcal{I} \ | \ r_{ui} = 1\}$ as a set of items interacted by user $u$. Let $\mathcal{D} = \{ (u,i) \ | \ u \in \mathcal{U} \wedge i \in I_u^+ \}$ denote a set of observed user-item pairs, and $\mathcal{T} = \{ (u,i,j) \ | \ u \in \mathcal{U} \wedge i \in \mathcal{I}_u^+ \wedge j \in \mathcal{I} \backslash \mathcal{I}_u^+ \}$ denote a set of triplets for positive and negative user-item pairs. Here, $\mathcal{T}$ is extended for a set of negative pairs. Let $\mathcal{T}_s = \{ (u, i, \mathcal{N}_u ) \ | \  u \in \mathcal{U} \wedge i \in \mathcal{I}_u^+ \wedge \mathcal{N}_u \subset \mathcal{I} \backslash \mathcal{I}_u^+ \}$ denote a set of triplets, where $\mathcal{N}_u$ is a set of negative items of user $u$. The goal of CF models is to infer a prediction score $\hat{r}_{ui}$ for unobserved user-item pair $(u, i)$ and recommend the items with the highest scores for user $u$.

\vspace{1mm}
\noindent
\textbf{Three key components of CF models}.
As discussed in existing work~\cite{MaoZWDDXH21}, the learning process of CF models consists of three parts, \ie, \emph{interaction encoder}, \emph{loss function}, and \emph{negative sampling}. (i) Interaction encoder: It learns user/item embeddings to capture collaborative signals. (ii) Loss function: CF models commonly utilize four types of loss functions such as \emph{pointwise}, \emph{pairwise}, \emph{setwise}, and \emph{alignment and uniformity (AU)} loss functions. Each loss function will be described in detail. (iii) Negative sampling: Because most entries in $\textbf{R}$ are unobserved, it is common to perform negative sampling for unobserved user-item pairs to improve training efficiency.

Although designing the interaction encoder is vital to capture complex and various correlations across users/items, recent studies~\cite{MaoZWDDXH21, WuWGCFQH22, GaoCPSV22, ZhangMWC22, WangYMZCLM22, DingQY0J20} report that loss functions and negative sampling are also beneficial to improve overall performance without designing complicated models. Notably, some studies~\cite{MaoZWDDXH21, WuWGCFQH22, WangYMZCLM22, ZhangMWC22} show that the simple MF model beats complicated neural models by simply modifying loss functions. Despite the recent progress, there is no formal study that investigates the relationship between various loss functions. This paper thus focuses on analyzing the relationship between various loss functions and proposes a novel loss function for CF models.

\subsection{Four Categories of Designing Loss Functions}

Table~\ref{tab:architecture_losses} depicts various loss functions for CF models. Each loss function expresses the relevance score between user $u$ and item $i$ as the cosine similarity or Euclidean distance. We denote the cosine similarity as $s(u,i)=\Tilde{f(u)}^{\top}\Tilde{f(i)}$ and Euclidean distance as $d(u,i)=||\Tilde{f(u)}-\Tilde{f(i)}||_2^2$, where $f(u)$ and $f(i) \in \mathbb{R}^{d}$ are user and item embedding vectors, respectively, and $\Tilde{f(\cdot)}$ means normalized representations. Existing loss functions are categorized as follows.
\begin{itemize}[leftmargin=5mm]
    \item \textbf{Pointwise loss function}: Because user-item pairs have either positive or negative values, it is natural to optimize them as the label of user-item pairs, \emph{e.g.}, BCE~\cite{HuKV08}, MCL~\cite{GaoCPSV22}, and UIB~\cite{ZhuoZYZ22}.
    \begin{align}
        \mathcal{L}_{Pointwise} = \sum_{(u, i) \in \mathcal{D}} \delta_{ui}
        + \sum_{(u, j) \notin \mathcal{D}} \delta_{uj}, \label{eq:pointwise}
    \end{align}
    where $\delta_{ui}$ and $\delta_{uj}$ denote a loss for positive pair $(u,i)$ and negative pair $(u,j)$, respectively. For BCE~\cite{HuKV08}, $\delta_{ui} = -\log \sigma (s(u,i))$ and $\delta_{uj} = - \log (1-\sigma (s(u,j)))$. 
    
    \vspace{1mm}
    \item \textbf{Pairwise loss function}: Because the pairwise loss function is designed to derive the maximum posterior estimator for personalized ranking, it is more suitable for addressing the top-$N$ recommendation problem. It includes BPR~\cite{RendleFGS09}, CML~\cite{HsiehYCLBE17}, and SML~\cite{LiZZQZHH20}. Because the number of triple pairs is proportional to the quadratic complexity for the number of items, it requires slow training time to converge.
    \begin{align}
        \mathcal{L}_{Pairwise} = \sum_{(u, i, j) \in \mathcal{T}} \delta_{uij}, \label{eq:pairwise}
    \end{align}
    where $\delta_{uij}$ means a loss for a triplet $(u,i,j)$. For BPR~\cite{RendleFGS09}, $\delta_{uij} = -\log \sigma (s(u,i) - s(u,j))$.
    
    \vspace{1mm}
    \item \textbf{Setwise loss function}: As the generalization of the pairwise loss function, it considers multiple negative item pairs at once. It includes CCL~\cite{MaoZWDDXH21}, SSM~\cite{WuWGCFQH22}, and BC~\cite{ZhangMWC22}. Since we typically sample $n$ negative items for each positive pair, it is more effective to consider a positive item and a set of negative items with equal weight and put them in one triplet.
    \begin{align}
        \mathcal{L}_{Setwise} = \sum_{(u, i, \mathcal{N}_{u}) \in \mathcal{T}_s} \delta_{ui\mathcal{N}_{u}}. \label{eq:setwise}
    \end{align}
    If a single negative item is used (\emph{i.e.}, $|\mathcal{N}_{u}| = 1$), it is equivalent to the pairwise loss function. For CCL~\cite{MaoZWDDXH21}, $\delta_{ui\mathcal{N}_{u}} = (1-s(u,i)) + \frac{w}{|\mathcal{N}_u|} \sum_{j \in \mathcal{N}_u} [s(u,j)-m]_+$, where $[\cdot]_+ = \max(0, \cdot)$.

    \vspace{1mm}
    \item \textbf{Alignment and uniformity loss function}:
    Recent studies~\cite{WangI20, GaoYC21, WangYMZCLM22} utilize the decomposition of the InfoNCE~\cite{abs-1807-03748} loss function into two losses, such as alignment and uniformity (AU) loss.
    
    \begin{align}
        \mathcal{L}_{AU} = \mathcal{L}_{Align} + \gamma \mathcal{L}_{Unif}, \label{eq:au}
    \end{align}
    
    where $\gamma$ denotes a hyperparameter to adjust alignment and uniformity. The alignment loss minimizes the distance between user $u$ and item $i$ for a positive pair $(u,i)$. This accomplishes the goal of CF, which is to recommend similar items to users with similar preferences.
    \begin{align}
        \mathcal{L}_{Align} = \displaystyle \mathop{\mathbb{E}}_{(u, i) \sim p_{pos}} d(u,i),
    \end{align}
    where $p_{pos}$ is the distribution for all positive user-item pairs. The uniformity loss makes all embedding vectors spread as far as possible on the hypersphere. This allows the hypersphere space to be maximally utilized to accommodate a variety of information. We compute the uniformity loss function on latent user and item spaces, respectively~\cite{WangYMZCLM22}.
    
    \begin{align}
        \mathcal{L}_{Unif} = \log \displaystyle \mathop{\mathbb{E}}_{\left(u, u^{\prime}\right) \sim p_{user}} e^{-2 \cdot d\left(u, u^{\prime}\right)}+\log \displaystyle \mathop{\mathbb{E}}_{\left(i, i^{\prime}\right) \sim p_{item}} e^{-2 \cdot d\left(i, i^{\prime}\right)},
    \end{align}
    where $p_{user}$ and $p_{item}$ represent the distribution for users and items, respectively. Besides, $u^{\prime}$ and $i^{\prime}$ are the in-batch user/item samples, respectively.

\end{itemize}

\section{Mathematical Analysis}\label{sec:analysis}

\begin{figure}
\includegraphics[height=5.1cm]{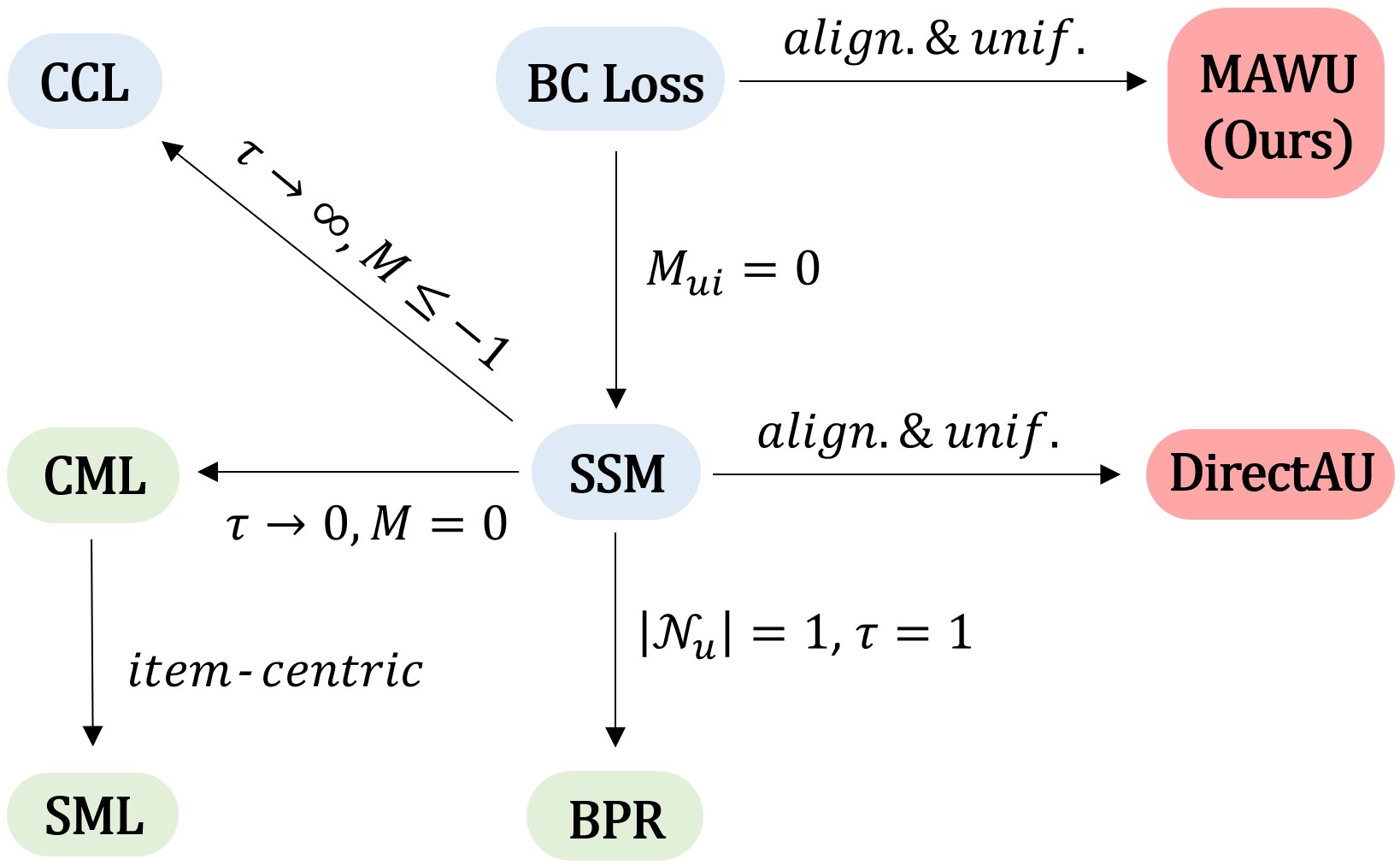} \\ 
\caption{Relationship diagram of CF loss functions except for the pointwise loss function. The \com{red}, \rev{blue}, and \green{green} backgrounds indicate alignment and uniformity, setwise, and pairwise loss functions, respectively.}\label{fig:relation_cf_losses}

\end{figure}

In this section, we investigate the relationships between various CF loss functions, with a fundamental setwise loss function, SSM~\cite{WuWGCFQH22}, as the reference. Figure~\ref{fig:relation_cf_losses} depicts the mathematical relationship across loss functions. Through our analysis, we reveal the following intriguing observations.

\begin{itemize}[leftmargin=5mm]
\item \textbf{Relationships between SSM~\cite{WuWGCFQH22} and pairwise loss functions}: Because setwise loss functions regard multiple negative samples at once, we derive three pairwise loss functions, \ie, CML~\cite{HsiehYCLBE17}, SML~\cite{LiZZQZHH20}, and BPR~\cite{RendleFGS09}, from SSM~\cite{WuWGCFQH22}.

\vspace{1mm}
\item \textbf{Relationships between SSM~\cite{WuWGCFQH22} and other setwise loss functions}: SSM~\cite{WuWGCFQH22} is induced to two setwise loss functions, \ie, CCL~\cite{MaoZWDDXH21} and BC~\cite{ZhangMWC22}. Besides, BC improves SSM using the margin for popularity bias.

\vspace{1mm}
\item \textbf{Relationship between SSM~\cite{WuWGCFQH22} and DirectAU~\cite{WangYMZCLM22}}: SSM~\cite{WuWGCFQH22} represents a contrastive loss, \ie, InfoNCE. Thus, DirectAU~\cite{WangYMZCLM22} using alignment and uniformity is derived from SSM. Because DirectAU does not require negative sampling, it learns better user/item representations for CF models than SSM.
\end{itemize}

Through our analysis, we found two potential improvements: (i) BC~\cite{ZhangMWC22} introduces a margin to adaptively optimize the user-item angular distance, and (ii) DirectAU~\cite{WangYMZCLM22} does not need any negative sampling strategies. However, there is no existing study that effectively combines the benefits of both loss functions.


\subsection{Analysis of Pairwise Loss Functions}

We first analyze the relationship between SSM~\cite{WuWGCFQH22} and CML~\cite{HsiehYCLBE17}. To derive CML~\cite{HsiehYCLBE17} from SSM~\cite{WuWGCFQH22}, we assume $\tau \rightarrow 0^{+}$, following the assumption of controlling $\tau$ in the InfoNCE loss in \cite{WangL21}. Here, $\tau$ is a temperature to control the skewness of the softmax distribution. When $\tau \rightarrow 0^{+}$, the SSM~\cite{WuWGCFQH22} is approximated to CML~\cite{HsiehYCLBE17} using the hardest negative samples. For simplicity, we focus on deriving a single triplet $\mathcal{T}_s$ $(u,i,\mathcal{N}_u)$ in Eq.~\eqref{eq:ssm}.
\begin{align}
    & \lim _{\tau \rightarrow \mathbf{0}^{+}} - \log \frac{e^{s(u,i) / \tau}}{e^{s(u,i) / \tau}+ \sum_{j \in \mathcal{N}_u} e^{s(u,j) / \tau}} \nonumber \\
    = & \lim _{\tau \rightarrow \mathbf{0}^{+}} + \log \left(1 + \sum_{j \in \mathcal{N}_u} e^{(s(u,j)-s(u,i)) / \tau}\right) \label{eq:lim_zero_2} \\
    \approx & \lim _{\tau \rightarrow \mathbf{0}^{+}} + \log \left(1 + \sum_{s(u,j) \ge s(u,i)}^{j} e^{(s(u,j)-s(u,i)) / \tau}\right) \label{eq:lim_zero_3} \\
    \approx & \lim _{\tau \rightarrow \mathbf{0}^{+}} \frac{1}{\tau} \left[s(u,j_{\max})-s(u, i)\right]_{+}, \label{eq:lim_zero_4}
\end{align}
where $j_{\max}$ is the negative item with the highest similarity with user $u$. As $\tau$ approaches zero in Eq.~\eqref{eq:lim_zero_2}, the gap between a positive pair and a negative pair gets larger. Thus, Eq.~\eqref{eq:lim_zero_2} is approximated by Eq.~\eqref{eq:lim_zero_3} regarding the negative pairs satisfying $s(u,j) \ge s(u,i)$. Finally, Eq.~\eqref{eq:lim_zero_3} is expressed by Eq.~\eqref{eq:lim_zero_4}, which considers only one pair with the hardest negative item $j_{\max}$.

Furthermore, CML~\cite{HsiehYCLBE17} has a margin. Assuming that the margin is zero, we finally derive the relationship between SSM and CML.
\begin{align}
    \lim _{\tau \rightarrow \mathbf{0}^{+}} \mathcal{L}_{SSM} & \approx \mathcal{L}_{CML} (m = 0) ~with~the~hardest~negative~item \nonumber \\
    & =\sum_{(u, i, j_{\max}) \in \mathcal{T}} \left[s(u,j_{\max})-s(u, i)\right]_{+} \label{eq:cml_sim} \\
    & \propto  \sum_{(u, i, j_{\max}) \in \mathcal{T}} \left[d(u, i)-d(u,j_{\max})\right]_{+}, \label{eq:cml_dist}     
\end{align}
where $s(u,i)=1-\frac{1}{2}(d(u,i))^2$ such that $\|d(u,i)\|_F = 1$. The similarity in Eq.~\eqref{eq:cml_sim} is thus converted to the Euclidean distance in Eq.~\eqref{eq:cml_dist}. Although SML~\cite{LiZZQZHH20} is different from CML~\cite{HsiehYCLBE17}, it is interpreted as combining both user- and item-centric loss functions. By deriving SSM~\cite{WuWGCFQH22} in both the user and item directions, it is possible to analyze the relationship between SSM~\cite{WuWGCFQH22} and SML~\cite{LiZZQZHH20}.

Lastly, we discuss how to derive BPR~\cite{RendleFGS09} from SSM~\cite{WuWGCFQH22}. When $|\mathcal{N}_u|= 1$ and $\tau=1$, it is equivalent to BPR~\cite{RendleFGS09}. Here, $\tau=1$ indicates that BPR does not employ hard negative filtering~\cite{WangL21}.
\begin{align}
    &\mathcal{L}_{SSM} (|\mathcal{N}_u|=1~\text{and}~\tau=1) = \mathcal{L}_{BPR} \nonumber \\
    & = - \sum_{(u, i, j) \in \mathcal{T}} \log \frac{e^{s(u,i)}}{e^{s(u,i)}+ e^{s(u,j)}} 
    = - \sum_{(u, i, j) \in \mathcal{T}} \log \sigma\left( s(u,i)-s(u,j) \right). \label{eq:bpr}
\end{align}

\subsection{Analysis of Setwise Loss Functions}
We analyze the relationship between SSM~\cite{WuWGCFQH22} and CCL~\cite{MaoZWDDXH21}. We employ the assumption of $\tau \rightarrow +\infty$ in \cite{WangL21}. When $\tau \rightarrow +\infty$, SSM is interpreted as CCL without hard negative filtering. 
\begin{align}
    & \lim _{\tau \rightarrow +\infty} - \log \frac{e^{s(u,i) / \tau}}{e^{s(u,i) / \tau}+ \sum_{j \in \mathcal{N}_u} e^{s(u,j) / \tau}} \nonumber \\
    = & \lim _{\tau \rightarrow +\infty} - \frac{1}{\tau} s(u,i) + \log \sum_{j \in \{\mathcal{N}_u, i\}} e^{s(u,j)/\tau} \label{eq:lim_inf_2} \\
    = & \lim _{\tau \rightarrow +\infty} - \frac{1}{\tau} s(u,i) + \log \left(\frac{1}{|\mathcal{N}_u|} \sum_{j \in \{\mathcal{N}_u, i\}} e^{s(u,j)/\tau} -1 +1 \right) + \log |\mathcal{N}_u| \label{eq:lim_inf_3} \\
    \approx & \lim _{\tau \rightarrow +\infty} - \frac{1}{\tau} s(u,i) + \frac{1}{|\mathcal{N}_u|} \sum_{j \in \{\mathcal{N}_u, i\}} e^{s(u,j)/\tau} -1 + \log |\mathcal{N}_u| \label{eq:lim_inf_4} \\
    \approx & \lim _{\tau \rightarrow +\infty} - \frac{|\mathcal{N}_u|-1}{|\mathcal{N}_u|\tau} s(u,i) + \frac{1}{|\mathcal{N}_u|\tau} \sum_{j \in \mathcal{N}_u} s(u,j) + \log |\mathcal{N}_u| + \frac{1}{|\mathcal{N}_u|} \label{eq:lim_inf_5} \\
    \approx & \lim _{\tau \rightarrow +\infty} - \frac{1}{\tau} s(u,i) + \frac{1}{|\mathcal{N}_u|\tau} \sum_{j \in \mathcal{N}_u} s(u,j) + \log |\mathcal{N}_u|. \label{eq:lim_inf_6}
\end{align}

Here, Eq.~\eqref{eq:lim_inf_5} is approximately derived by utilizing Taylor expansion of $\log(1+x)$ and $\exp(x)$ in Eq.~\eqref{eq:lim_inf_3} and Eq.~\eqref{eq:lim_inf_4}, respectively, \ie, $\log (1+x) \approx x$ and $\exp (x) \approx 1+x$, for $|x| \ll 1$. As $\tau$ approaches infinity, $(\frac{1}{|\mathcal{N}_u|} \sum_{j \in \{\mathcal{N}_u, i\}} e^{s(u,j)/\tau} -1)$ in Eq.~\eqref{eq:lim_inf_3} and $s(u,j)/\tau$ in Eq.~\eqref{eq:lim_inf_4} reach very small values. Thus, when using Taylor expansion, the second- and higher-order terms can be ignored. Finally, because $|\mathcal{N}_u| \gg 1$ in Eq.~\eqref{eq:lim_inf_5}, it is approximated by Eq.~\eqref{eq:lim_inf_6}.

We explain how to derive CCL~\cite{MaoZWDDXH21} from SSM~\cite{WuWGCFQH22}. It has two heuristic hyperparameters for the weight $w$ of negative samples and the margin $M$ for hard negative filtering. Assuming $w = 1$ and no negative filtering, CCL~\cite{MaoZWDDXH21} is interpreted by $\lim_{\tau \rightarrow +\infty} \mathcal{L}_{SSM}$.
\begin{align}
    \lim _{\tau \rightarrow +\infty} \mathcal{L}_{SSM} & \approx \mathcal{L}_{CCL}(w=1)~without~hard~negative~filtering \nonumber \\
    & =\sum_{(u,i,\mathcal{N}_u) \in \mathcal{T}_s}\left(1-s(u,i) + \frac{1}{|\mathcal{N}_u|} \sum_{j \in \mathcal{N}_u} s(u,j)\right) \label{eq:simple}.
\end{align}

We then deal with the relationship between SSM~\cite{WuWGCFQH22} and BC~\cite{ZhangMWC22}. While both loss functions employ the softmax loss function, BC also uses a margin. Assuming $M_{ui}=0$, BC is induced to SSM. SSM does not adjust weight based on positive pairs, and it is difficult to train hard positive samples, which may degrade the model performance.
\begin{align}
    \mathcal{L}_{BC}(M_{ui}=0) & = \mathcal{L}_{SSM} \nonumber \\
    & = - \sum_{(u,i,\mathcal{N}_u) \in \mathcal{T}_s} \log \frac{e^{s(u,i) / \tau}}{e^{s(u,i) / \tau}+ \sum_{j \in \mathcal{N}_u} e^{s(u,j) / \tau}}, \label{eq:ssm}
\end{align}
where $M_{ui}$ is an interaction-wise margin for user $u$ and item $i$.

\subsection{Analysis of Alignment and Uniformity}

We derive the relationship between SSM~\cite{WuWGCFQH22} and DirectAU~\cite{WangYMZCLM22}. According to Theorem 1 in \cite{WangI20}, InfoNCE, \ie, SSM~\cite{WuWGCFQH22} is understood by optimizing alignment and uniformity. Specifically, for fixed $\tau > 0$ and $|\mathcal{N}_u| \rightarrow \infty$, it is also proved that SSM asymptotically converges to the combination of alignment and uniformity loss.
\begin{align}
  & \lim_{|\mathcal{N}_u| \rightarrow \infty} - \displaystyle \mathop{\mathbb{E}}_{(u, i) \sim p_{pos}} \left [ \log \frac{e^{s(u,i) / \tau} }{e^{s(u,i) / \tau} + \sum_{j \in \mathcal{N}_u} e^{s(u,j) / \tau}} \right ] - \log |\mathcal{N}_u| \label{eq:logarithm} \\
  = & -\frac{1}{\tau} \displaystyle \mathop{\mathbb{E}}_{(u, i) \sim p_{pos}} s(u,i) \nonumber \\
  + & \displaystyle \mathop{\mathbb{E}}_{(u, i) \sim p_{pos}} \left[\lim_{|\mathcal{N}_u| \rightarrow \infty} \log \left(\frac{1}{|\mathcal{N}_u|} e^{s(u,i) / \tau} + \frac{1}{|\mathcal{N}_u|} \sum_{j \in \mathcal{N}_u} e^{s(u,j) / \tau}\right)\right] \label{eq:num_neg_larger} \\
  = & -\frac{1}{\tau} \displaystyle \mathop{\mathbb{E}}_{(u,i) \sim p_{pos}} s(u,i) + \displaystyle \mathop{\mathbb{E}}_{u \sim p_{data}} \left[ \log \displaystyle \mathop{\mathbb{E}}_{j \sim p_{data}} e^{s(u,j) / \tau} \right]\label{eq:uniformity} \\
  \approx & -\frac{1}{\tau} \displaystyle \mathop{\mathbb{E}}_{(u, i) \sim p_{pos}} s(u,i) \nonumber \\
  + & \log \displaystyle \mathop{\mathbb{E}}_{\left(u, u^{\prime}\right) \sim p_{user}} e^{-2 \cdot d\left(u, u^{\prime}\right)}+\log \displaystyle \mathop{\mathbb{E}}_{\left(i, i^{\prime}\right) \sim p_{item}} e^{-2 \cdot d\left(i, i^{\prime}\right)}. \label{eq:ssm_au}
\end{align}
Here, $p_{data}$ denotes the distribution of data. We derive Eq.~\eqref{eq:uniformity} from Eq.~\eqref{eq:num_neg_larger} using the strong law of large numbers and the continuous mapping theorem. Then, the second term of Eq.~\eqref{eq:uniformity} is divided by the user and item uniformity, using the logarithm of average pairwise Gaussian potential in~\cite{WangI20}.

Surprisingly, Eq.~\eqref{eq:ssm_au} is the equivalent form to DirectAU~\cite{WangYMZCLM22} using alignment and uniformity for recommendation. SSM~\cite{WuWGCFQH22} is asymptotically approximated by combining the alignment of positive pairs and the uniformity for users and items. Besides, the temperature $\tau$ in Eq.~\eqref{eq:ssm_au} is interpreted as a hyperparameter to control the significance between alignment and uniformity, as used in $\gamma$ in Eq~\eqref{eq:au}. As $\tau$ increases, it highlights the uniformity.

From the previous analysis of setwise and pairwise loss functions, most of them are distinguished by the negative sampling-related parameters (\ie, temperature $\tau$ and the number of negatives $|\mathcal{N}_u|$). Since DirectAU~\cite{WangYMZCLM22} is SSM~\cite{WuWGCFQH22} as $|\mathcal{N}_u|$ goes infinity, DirectAU can have the same effect as sampling many negative items in SSM. Thus, DirectAU achieves the effect of negative sampling without any sampling strategies, leading to better performance than SSM.


\section{Proposed Loss Function}\label{sec:model}


In this section, we propose a novel loss function by considering the unique characteristics of a target dataset, called \emph{Margin-aware Alignment and Weighted Uniformity (MAWU)}. It takes the advantages of both BC~\cite{ZhangMWC22} and DirectAU~\cite{WangYMZCLM22}. Our proposed MAWU is composed of two loss functions: (i) \emph{margin-aware alignment (MA)} incorporates the margin into the alignment loss, and (ii) \emph{weighted uniformity (WU)} controls the importance of users and items depending on user/item distributions.


\vspace{1mm}
\noindent
\textbf{Margin-aware alignment loss.} Our analysis revealed that (i) BC~\cite{ZhangMWC22} is interpreted as SSM~\cite{WuWGCFQH22} with the margin. (ii) DirectAU~\cite{WangYMZCLM22} directly learns user/item representations without negative sampling. Based on the relationship between DirectAU and SSM, we first transform BC into a combination of alignment and uniformity.
\begin{align}
  & \lim _{|\mathcal{N}_u| \rightarrow +\infty} - \displaystyle \mathop{\mathbb{E}}_{(u, i) \sim p_{pos}} \left[ \log \frac{e^{\cos \left(\hat{\theta}_{u i}+M_{u i}\right) / \tau} }{e^{\cos \left(\hat{\theta}_{u i}+M_{u i}\right) / \tau} + \sum_{j \in \mathcal{N}_u} e^{\cos \left(\hat{\theta}_{u j}\right) / \tau}} \right] \nonumber \\
  & - \log |\mathcal{N}_u| \label{eq:bcloss} \\
  & \approx -\frac{1}{\tau} \displaystyle \mathop{\mathbb{E}}_{(u, i) \sim p_{pos}} \cos \left(\hat{\theta}_{u i}+M_{u i}\right) \nonumber \\
  & + \log \displaystyle \mathop{\mathbb{E}}_{\left(u, u^{\prime}\right) \sim p_{user}} e^{-2 \cdot d\left(u, u^{\prime}\right)}+\log \displaystyle \mathop{\mathbb{E}}_{\left(i, i^{\prime}\right) \sim p_{item}} e^{-2 \cdot d\left(i, i^{\prime}\right)}. \label{eq:bcloss_au}
\end{align}
Here, $\hat{\theta}_{ui}$ means the angle between $f(u)$ and $f(i)$. The margin in BC~\cite{ZhangMWC22} is used to discriminate the gap between positive pair $(u, i)$ and negative pair $(u, j)$, \ie, $\cos(\hat{\theta}_{ui} + M_{ui}) > \cos(\hat{\theta}_{uj})$, penalizing the angular distance between user $u$ and item $i$ in positive pair $(u, i)$. In the alignment loss, we also employ the margin to make a large loss value, stretching a user-item angular distance.

\begin{figure}

\centering
\begin{tabular}{cc}
\renewcommand{\arraystretch}{0.9} 
\includegraphics[width=0.19\textwidth]{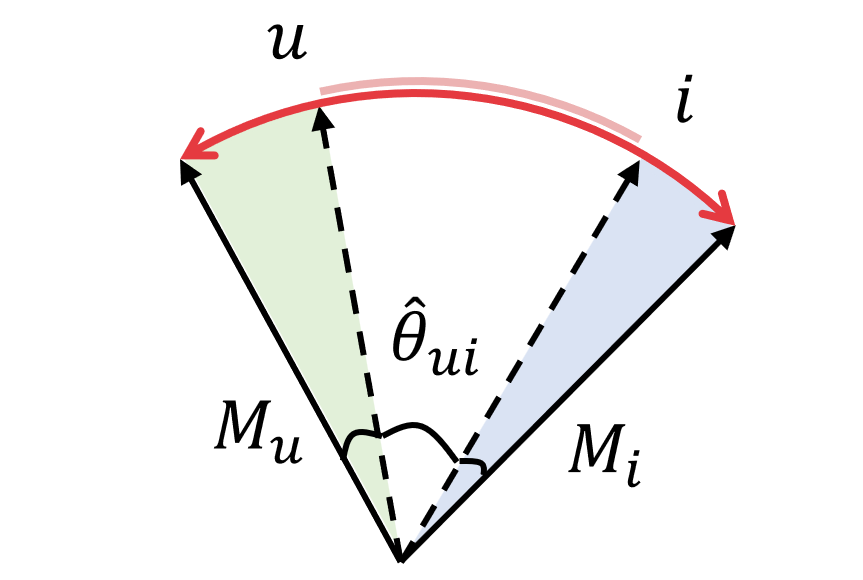} &
\includegraphics[width=0.19\textwidth]{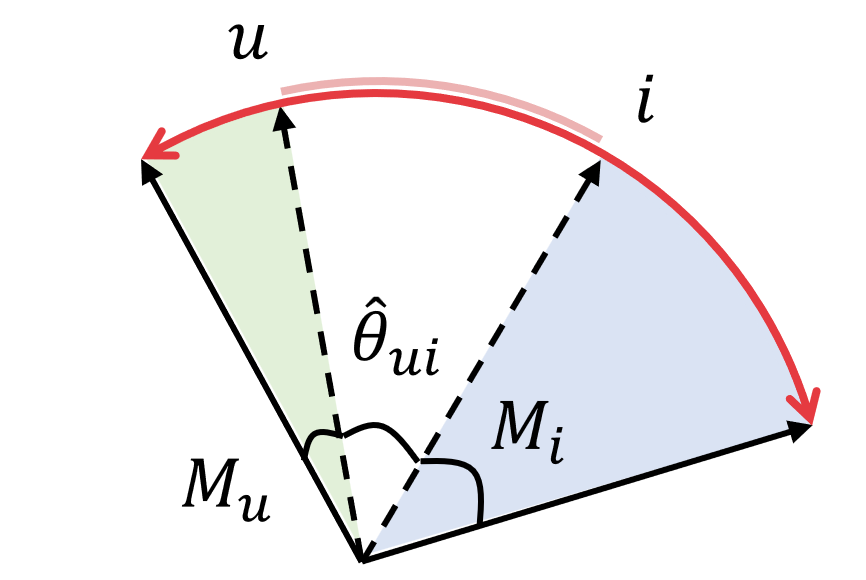} \\
\multicolumn{1}{c}{(a) Pop. $u$ and pop. $i$} &\multicolumn{1}{c}{(b) Pop. $u$ and unpop. $i$} \vspace{1.5mm} \\
\includegraphics[width=0.19\textwidth]{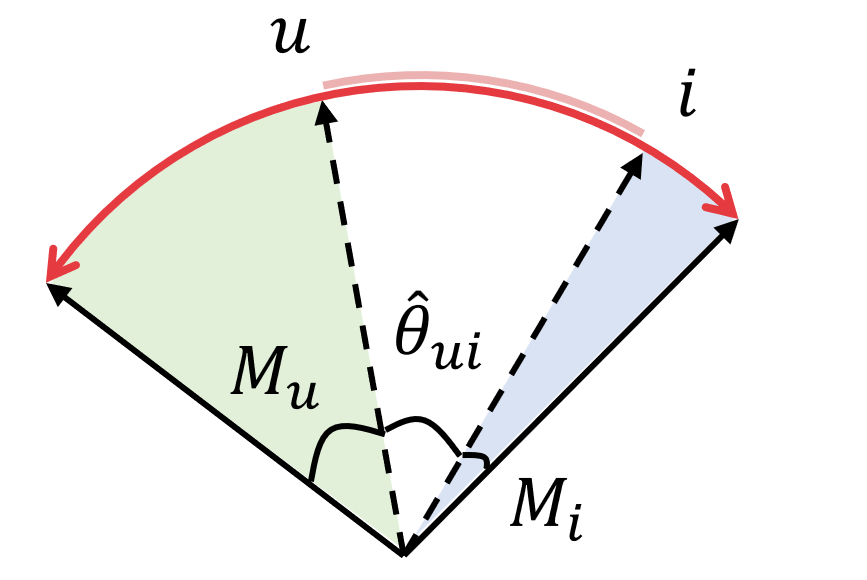} &
\includegraphics[width=0.19\textwidth]{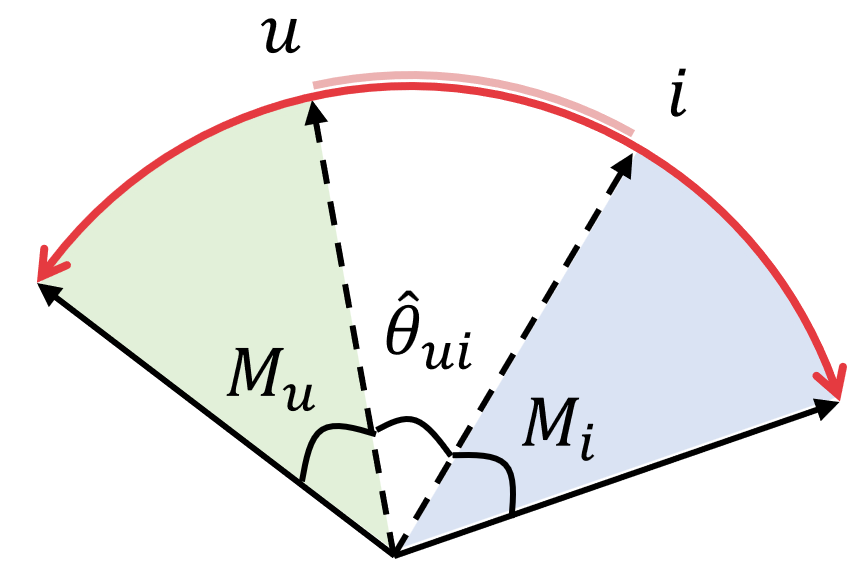} \\
\multicolumn{1}{c}{(c) Unpop. $u$ and pop. $i$} &\multicolumn{1}{c}{(d) Unpop. $u$ and unpop. $i$}
\vspace{-2mm}
\end{tabular}

\caption{Four user $u$-item $i$ cases in the training process for margin-aware alignment loss of MAWU. As training progresses, unpopular users and items have larger margin values than popular users and items. The \green{green} and \rev{blue} backgrounds mean user and item margins, respectively.}\label{fig:proposed_method}

\end{figure}



Specifically, we devise a new margin function that combines two margins for users and items. Because the margin varies depending on users and items, we represent the margin as a sum of a user margin and an item margin, \ie, $M_{ui} = M_{u} + M_{i}$, where $M_{u} \in \mathbb{R}^{|\mathcal{U}|}$ and $M_{i} \in \mathbb{R}^{|\mathcal{I}|}$ are trainable parameters. While BC~\cite{ZhangMWC22} additionally introduces a popularity encoder to estimate $M_{ui}\in \mathbb{R}^{|\mathcal{U}| \times |\mathcal{I}|}$, we simply use $|\mathcal{U}|+|\mathcal{I}|$ parameters without an additional module. In Appendix~\ref{appen:proof_ma}, we show theoretically that MA loss function can learn more user/item-specific embeddings than BC~\cite{ZhangMWC22}.

Despite its simplicity, it can capture the user/item-specific margins and mitigate the biases from the active users and popular items.
Figure~\ref{fig:proposed_method} depicts four user-item cases in the training process based on user/item popularity. Particularly, it emphasizes positive pairs with unpopular users and items (See Figure~\ref{fig:proposed_method}(d)). It is effective for adaptively learning them with the margin depending on user and item popularity. Because the cosine function is non-monotonic, we also restrict the range of the cosine function to make it monotonic, \ie, $0 \le \hat{\theta}_{ui} + M_{u} + M_{i} \le \pi$. Formally, we formulate the margin-aware alignment loss function.
\begin{align}\label{eq:m_align}
  \mathcal{L}_{MA} = - \displaystyle \mathop{\mathbb{E}}_{(u, i) \sim p_{pos}} \cos \left(\hat{\theta}_{u i}+M_u+M_i\right).
\end{align}

\vspace{1mm}
\noindent
\textbf{Weighted uniformity loss.} Existing studies~\cite{WangI20, WangYMZCLM22} have used the same weights for user and item uniformity loss functions without considering the difference between users and items. However, in real-world recommendation datasets, the distribution of interactions can vary depending on users and items. If this perspective is neglected, it may lead to sub-optimal results. Table~\ref{tab:statistics} provides the different ratios of Gini indices~\cite{ChinCC22} between users and items in three datasets. Here, $Gini_u$ and $Gini_i$ represent the Gini indices that measure the distribution of interactions within the sets of users $\mathcal{U}$ and items $\mathcal{I}$, respectively. A small $Gini_u$ means that most users tend to rate items evenly. Therefore, it is necessary to highlight user uniformity if the dataset has more diversity in users than items, \ie, the dataset has high $Gini_i/Gini_u$.

Based on these observations, we adopt different weights for user and item uniformities for each dataset. While this approach introduces additional hyperparameters, it enables us to more effectively address the statistical differences between users and items. Formally, we represent the weighted uniformity loss function as follows.

\begin{align}\label{eq:w_unif}
  \mathcal{L}_{WU} & = \gamma_1 \log \displaystyle \mathop{\mathbb{E}}_{\left(u, u^{\prime}\right) \sim p_{user}} e^{-2 \cdot d\left(u, u^{\prime}\right)} + \gamma_2 \log \displaystyle \mathop{\mathbb{E}}_{\left(i, i^{\prime}\right) \sim p_{item}} e^{-2 \cdot d\left(i, i^{\prime}\right)}, 
\end{align}
where $\gamma_1$ and $\gamma_2$ are the hyperparameters to control user and item uniformities, respectively. 

\vspace{1mm}
\noindent
\textbf{Training and Inference.} For model training, we lastly formulate the MAWU loss function that combines margin-aware alignment and weighted uniformity loss functions.
\begin{align}
  \mathcal{L}_{MAWU} = \mathcal{L}_{MA} + \mathcal{L}_{WU}.
\end{align}

For inference, we make predictions using the inner product of user and item embedding vectors without normalization. Note that the user and item margins are only used for robust model training.
\begin{align}\label{eq:pred}
  \hat{r}_{ui} = f(u)^{\top}f(i).
\end{align}
\makeatletter
\def\thickhline{%
  \noalign{\ifnum0=`}\fi\hrule \@height \thickarrayrulewidth \futurelet
   \reserved@a\@xthickhline}
\def\@xthickhline{\ifx\reserved@a\thickhline
               \vskip\doublerulesep
               \vskip-\thickarrayrulewidth
             \fi
      \ifnum0=`{\fi}}
\makeatother

\newlength{\thickarrayrulewidth}
\setlength{\thickarrayrulewidth}{1.5\arrayrulewidth}

\begin{table}[t] 
\centering
\caption{Statistics of the datasets after preprocessing and the ratios of Gini indices between users (\ie, $Gini_{u}$) and items (\ie, $Gini_{i}$) for each dataset.}\label{tab:statistics}
\vspace{-4mm}
\begin{center}
\renewcommand{\arraystretch}{0.95} 
\begin{tabular}{P{1.2cm}P{0.8cm}P{0.8cm}P{1.0cm}P{0.9cm}P{1.2cm}}
\toprule
\multicolumn{1}{c}{Dataset} & \multicolumn{1}{c}{\#user} & \multicolumn{1}{c}{\#item} & \multicolumn{1}{c}{\#inter.}  & \multicolumn{1}{c}{Density} & \multicolumn{1}{c}{$Gini_{i}/Gini_{u}$} \\
\hline
Beauty & 22.4k & 12.1k & 198.5k & 0.07\% & 2.096 \\
Gowalla & 29.9k & 41.0k & 1,027.4k & 0.08\% & 0.518 \\
Yelp2018 & 31.7k & 38.0k & 1,561.4k & 0.13\% & 1.090 \\

\bottomrule
\end{tabular}
\end{center}
\vspace{-3mm}
\end{table}

\vspace{-1mm}
\vspace{-1mm}
\section{Experimental Setup}
\label{sec:setup}

\noindent
\textbf{Datasets and preprocessing}. We used three benchmark datasets, Beauty\footnote{\vspace{-0.5mm}\url{https://jmcauley.ucsd.edu/data/amazon/links.html}}, Gowalla\footnote{\vspace{-0.5mm}\url{http://snap.stanford.edu/data/loc-gowalla.html}}, and Yelp2018\footnote{\vspace{-0.5mm}\url{https://www.yelp.com/dataset}}, widely used in existing studies~\cite{WangYMZCLM22, 0001DWLZ020, MaoZWDDXH21}. Following~\cite{WangYMZCLM22, Wang0WFC19}, we use the 5-core setting for Beauty and the 10-core setting for the other datasets. Also, we leave the last interacted item if item interactions are duplicated. Table~\ref{tab:statistics} reports the statistics of the three datasets after preprocessing. We split each dataset into train:valid:test sets with a ratio of 7:1:2.

\begin{table*}[t] \small
\centering
\caption{Accuracy comparison of ours and baselines on Beauty, Gowalla, and Yelp2018. $\dagger$ indicates the best baseline, \ie, DAU for t-test and performance gain. * and ** indicate $p < 0.05$ and $p < 0.01$ for a one-tailed t-test. Bold indicates the best performance between MAWU and the best baseline, \ie, DAU. $\ddagger$ indicates higher performance than MAWU\protect\footnotemark.}\label{tab:cf_losses}

\vspace{-4mm}
\begin{center}
\renewcommand{\arraystretch}{1} 
\newcommand\Tstrut{\rule{0pt}{2.3ex}}         
\begin{tabular} {P{1.0cm}P{1.65cm}P{0.85cm}P{0.85cm}P{0.85cm}P{0.85cm}P{0.85cm}P{0.85cm}P{0.85cm}P{0.85cm}P{0.85cm}P{0.85cm}P{0.85cm}P{0.85cm}P{0.85cm}}
\toprule

\multicolumn{2}{c}{} & \multicolumn{6}{c}{Backbone: MF} & \multicolumn{6}{c}{Backbone: LightGCN} \\
\cmidrule(lr){3-8}
\cmidrule(lr){9-14}

\multicolumn{2}{c}{} & \multicolumn{3}{c}{Recall$@N$} &  \multicolumn{3}{c}{NDCG$@N$} & \multicolumn{3}{c}{Recall$@N$} &  \multicolumn{3}{c}{NDCG$@N$}\\
\cmidrule(lr){3-5}
\cmidrule(lr){6-8}
\cmidrule(lr){9-11}
\cmidrule(lr){12-14}
 Dataset & Loss & $N$=10 & $N$=20 & $N$=50 & $N$=10 & $N$=20 & $N$=50 & $N$=10 & $N$=20 & $N$=50 & $N$=10 & $N$=20 & $N$=50 \\
\hline
\multirow{12}{*}{Beauty}\Tstrut
&  BCE~\cite{HuKV08} & 0.0738 & 0.1065 & 0.1618 & 0.0450 & 0.0542 & 0.0666 & 0.0920 & 0.1304 & 0.1953 & 0.0555 & 0.0663 & 0.0808 \\
 & MCL~\cite{GaoCPSV22} & 0.0922 & 0.1304 & 0.1944 & 0.0569 & 0.0676 & 0.0819 & 0.0892 & 0.1284 & 0.1947 & 0.0535 & 0.0644 & 0.0793 \\
 & UIB~\cite{ZhuoZYZ22} & 0.0829 & 0.1189 & 0.1769 & 0.0499 & 0.0600 & 0.0731 & 0.0895 & 0.1270 & 0.1883 & 0.0541 & 0.0647 & 0.0784 \\
  \cline{2-14}\Tstrut
 & BPR~\cite{RendleFGS09} & 0.0885 & 0.1271 & 0.1944 & 0.0533 & 0.0641 & 0.0791 & 0.0882 & 0.1283 & 0.1992 & 0.0521 & 0.0632 & 0.0789 \\
 & CML~\cite{HsiehYCLBE17} & 0.0545 & 0.0828 & 0.1326 & 0.0329 & 0.0410 & 0.0523 & 0.0328 & 0.0492 & 0.0817 & 0.0202 & 0.0249 & 0.0325 \\
 & SML~\cite{LiZZQZHH20} & 0.0712 & 0.1055 & 0.1645 & 0.0422 & 0.0518 & 0.0652 & 0.0748 & 0.1119 & 0.1756 & 0.0442 & 0.0545 & 0.0686 \\
  \cline{2-14}\Tstrut
 & CCL~\cite{MaoZWDDXH21} &0.0751 & 0.1036 & 0.1489 & 0.0483 & 0.0563 & 0.0664 & 0.0716 & 0.0980 & 0.1394 & 0.0451 & 0.0524 & 0.0617 \\
 & SSM~\cite{WuWGCFQH22} & 0.0891 & 0.1314 & 0.1986 & 0.0537 & 0.0655 & 0.0805 & 0.0870 & 0.1261 & 0.1958 & 0.0517 & 0.0627 & 0.0782 \\
 & BC~\cite{ZhangMWC22} & 0.0877 & 0.1274 & 0.1911 & 0.0527 & 0.0639 & 0.0782 & 0.0874 & 0.1276 & 0.1961 & 0.0522 & 0.0633 & 0.0786 \\
\cline{2-14}\Tstrut
& DAU$^{\dagger}$~\cite{WangYMZCLM22} & 0.0975 & 0.1399 & \textbf{0.2064}$^{\ddagger}$ & 0.0598 & 0.0716 & 0.0865 & 0.0983 & 0.1412 & 0.2134 & 0.0592 & 0.0711 & 0.0872 \\
\cline{2-14}\Tstrut
& MAWU (ours) & \textbf{0.1004*} & \textbf{0.1405} & 0.2036 & \textbf{0.0628**} & \textbf{0.0740**} & \textbf{0.0881*} & \textbf{0.1030*} & \textbf{0.1457*} & \textbf{0.2135} & \textbf{0.0635**} & \textbf{0.0754**} & \textbf{0.0905**} \\
\cline{2-14}
& Gain (\%) & 3.02 & 0.40 & -1.34 & 5.05 & 3.35 & 1.90 & 4.74 & 3.16 & 0.07 & 7.30 & 6.10 & 3.83 \\
\hline
\multirow{12}{*}{Gowalla}\Tstrut
& BCE~\cite{HuKV08} & 0.1181 & 0.1731 & 0.2763 & 0.1132 & 0.1302 & 0.1614 & 0.1343 & 0.1938 & 0.3027 & 	0.1307 & 0.1482 & 0.1807 \\
 & MCL~\cite{GaoCPSV22} & 0.1319 & 0.1946$^{\ddagger}$ & 0.3085$^{\ddagger}$ & 0.1242 & 0.1433 & 0.1775 & 0.1122 & 0.1635 & 0.2628 & 0.1092 & 0.1245 & 0.1543 \\
 & UIB~\cite{ZhuoZYZ22} & 0.1177 & 0.1755 & 0.2838 & 0.1112 & 0.1293 & 0.1619 & 0.1295 & 0.1877 & 0.2952 & 0.1251 & 0.1425 & 0.1747 \\
  \cline{2-14}\Tstrut
 & BPR~\cite{RendleFGS09} & 0.1153 & 0.1692 & 0.2713 & 0.1084 & 0.1255 & 0.1562 & 0.1284 & 0.1866 & 0.2952 & 0.1247 & 0.1422 & 0.1747 \\
 & CML~\cite{HsiehYCLBE17} & 0.0590 & 0.0987 & 0.1848 & 0.0519 & 0.0652 & 0.0915 & 0.0485 & 0.0810 & 0.1515 & 0.0430 & 0.0538 & 0.0752 \\
 & SML~\cite{LiZZQZHH20} & 0.0710 & 0.1169 & 0.2130 & 0.0623 & 0.0775 & 0.1066 & 0.0559 & 0.0902 & 0.1611 & 0.0458 & 0.0580 & 0.0800 \\
  \cline{2-14}\Tstrut
 & CCL~\cite{MaoZWDDXH21} & 0.1394$^{\ddagger}$ & 0.1993$^{\ddagger}$ & 0.3003 & 0.1377$^{\ddagger}$ & 0.1543$^{\ddagger}$ & 0.1838$^{\ddagger}$ & 0.1284 & 0.1847 & 0.2794 & 0.1264 & 0.1422 & 0.1698 \\
 & SSM~\cite{WuWGCFQH22} & 0.1215 & 0.1819 & 0.2940 & 0.1137 & 0.1324 & 0.1661 & 0.1093 & 0.1601 & 0.2609 & 0.1059 & 0.1212 & 0.1514 \\
 & BC~\cite{ZhangMWC22} & 0.1194 & 0.1821 & 0.3007 & 0.1104 & 0.1300 & 0.1655 & 0.1094 & 0.1603 & 0.2605 & 0.1061 & 0.1213 & 0.1514 \\
\cline{2-14}\Tstrut
& DAU$^{\dagger}$~\cite{WangYMZCLM22} & 0.1293 & 0.1881 & 0.2979 & 0.1257 & 0.1431 & 0.1758 & 0.1313 & 0.1905 & 0.3010 & 0.1287 & 0.1461 & 0.1789 \\
\cline{2-14}\Tstrut
& MAWU (ours) & \textbf{0.1326}** & \textbf{0.1921}** & \textbf{0.3022}** & \textbf{0.1277}** & \textbf{0.1454}** & \textbf{0.1783}** & \textbf{0.1351}** & \textbf{0.1956}** & \textbf{0.3074}** & \textbf{0.1317}** & \textbf{0.1496}** & \textbf{0.1829}** \\
\cline{2-14}
& {Gain (\%)} & 2.55 & 2.08 & 1.45 & 1.58 & 1.59 & 1.40 & 2.86 & 2.68 & 2.11 & 2.35 & 2.41 & 2.24 \\
\hline
\multirow{12}{*}{Yelp2018}\Tstrut
& BCE~\cite{HuKV08} & 0.0532 & 0.0879 & 0.1615 & 0.0606 & 0.0723 & 0.0984 & 0.0594 & 0.0973 & 0.1772 & 0.0691 & 0.0814 & 0.1094 \\
 & MCL~\cite{GaoCPSV22} & 0.0608 & 0.1000 & 0.1819 & 0.0705 & 0.0831 & 0.1118 & 0.0542 & 0.0895 & 0.1647 & 0.0624 & 0.0742 & 0.1007 \\
 & UIB~\cite{ZhuoZYZ22} & 0.0468 & 0.0791 & 0.1500 & 0.0535 & 0.0644 & 0.0897 & 0.0582 & 0.0956 & 0.1745 & 0.0669 & 0.0792 & 0.1070 \\
 \cline{2-14}\Tstrut
 & BPR~\cite{RendleFGS09} & 0.0434 & 0.0729 & 0.1384 & 0.0486 & 0.0588 & 0.0820 & 0.0564 & 0.0921 & 0.1693 & 0.0647 & 0.0764 & 0.1035 \\
 & CML~\cite{HsiehYCLBE17} & 0.0292 & 0.0519 & 0.1070 & 0.0333 & 0.0414 & 0.0612 & 0.0400 & 0.0698 & 0.1379 & 0.0455 & 0.0558 & 0.0801 \\
 & SML~\cite{LiZZQZHH20} & 0.0352 & 0.0614 & 0.1227 & 0.0399 & 0.0491 & 0.0709 & 0.0295 & 0.0524 & 0.1075 & 0.0334 & 0.0415 & 0.0614 \\
 \cline{2-14}\Tstrut
 & CCL~\cite{MaoZWDDXH21} & 0.0521 & 0.0823 & 0.1415 & 0.0614 & 0.0706 & 0.0910 & 0.0522 & 0.0831 & 0.1459 & 0.0613 & 0.0708 & 0.0927 \\
 & SSM~\cite{WuWGCFQH22} & 0.0566 & 0.0937 & 0.1722 & 0.0657 & 0.0777 & 0.1052 & 0.0486 & 0.0809 & 0.1513 & 0.0564 & 0.0671 & 0.0919 \\
 & BC~\cite{ZhangMWC22} & 0.0606 & 0.1010 & 0.1856 & 0.0699 & 0.0831 & 0.1128 & 0.0492 & 0.0815 & 0.1529 & 0.0568 & 0.0676 & 0.0928 \\
 \cline{2-14}\Tstrut
 & DAU$^{\dagger}$~\cite{WangYMZCLM22} & 0.0639 & 0.1031 & 0.1834 & 0.0745 & 0.0870 & 0.1151 & 0.0658 & 0.1054 & 0.1872 & 0.0773 & 0.0897 & 0.1182 \\
\cline{2-14}\Tstrut
& MAWU (ours) & \textbf{0.0656**} & \textbf{0.1058**} & \textbf{0.1871**} & \textbf{0.0766**} & \textbf{0.0894**} & \textbf{0.1178**} & \textbf{0.0664} & \textbf{0.1066**} & \textbf{0.1895**} & \textbf{0.0779*} & \textbf{0.0905*} & \textbf{0.1194**} \\
\cline{2-14}
& Gain (\%) & 2.66 & 2.62 & 2.02 & 2.82 & 2.76 & 2.35 & 0.85 & 1.14 & 1.25 & 0.80 & 0.89 & 1.00 \\
\bottomrule
\end{tabular}
\end{center}
\end{table*}

\footnotetext{\vspace{-0.4mm}On Gowalla-MF, CCL has salient performance because Gowalla has a relatively uniform distribution of items, and CCL highlights more uniformity ($\tau$ goes to infinity).}

\vspace{1mm}
\noindent
\textbf{Baseline loss functions and CF models}. We compare MAWU with ten CF loss functions in Table~\ref{tab:architecture_losses}. To compare different loss functions, we used MF~\cite{HuKV08} and LightGCN~\cite{0001DWLZ020} as the backbone model. Specifically, ten loss functions used in the evaluation are categorized into four groups: (i) Pointwise loss (\ie, BCE~\cite{HuKV08}, MCL~\cite{GaoCPSV22}, and UIB~\cite{ZhuoZYZ22}) (ii) Pairwise loss (\ie, BPR~\cite{RendleFGS09}, CML~\cite{HsiehYCLBE17}, and SML~\cite{LiZZQZHH20}) (iii) Setwise loss (\ie, CCL~\cite{MaoZWDDXH21}, SSM~\cite{WuWGCFQH22}, and BC Loss~\cite{ZhangMWC22}) (iv) DirectAU~\cite{WangYMZCLM22}. To further verify the competitive edge of our loss function, it is compared with eight state-of-the-art CF models. We adopt three MF-based models (\ie, NeuMF~\cite{HeLZNHC17}, ENMF~\cite{ChenZZLM20}, and SimpleX~\cite{MaoZWDDXH21}) and two AE-based models (\ie, MultVAE~\cite{LiangKHJ18} and RecVAE~\cite{ShenbinATMN20}) and two GNN-based models (\ie, SGL~\cite{WuWF0CLX21}, SimGCL~\cite{YuY00CN22}, and LightGCL~\cite{CaiHXR23}). To verify the effectiveness of only the loss function, random sampling is used in MCL~\cite{GaoCPSV22}, and metadata (\eg, item features) is not used in CML~\cite{HsiehYCLBE17}. Besides, UIB~\cite{ZhuoZYZ22} leverages BPR~\cite{RendleFGS09} as a backbone loss function.


\vspace{1mm}
\noindent
\textbf{Evaluation metrics}. We use two popular metrics, \emph{Normalized Discounted Cumulative Gain (NDCG@$N$)} and \emph{Recall@$N$}, to evaluate \topN\ recommendation. We use $N = \{10, 20, 50\}$ for all datasets.

\vspace{1mm}
\noindent
\textbf{Reproducibility}. All the methods are implemented in Recbole~\cite{ZhaoMHLCPLLWTMF21} framework and optimized using Adam~\cite{KingmaB14} optimizer, and their parameters are initialized with Xavier's method~\cite{GlorotB10}. For all CF models, we set latent dimension, learning rate, batch size, and the number of negative samples to 64, 0.001, 2,048, and 30, respectively. Besides, weight decay is tuned in \{0, 1e-2, 1e-4, 1e-6, 1e-8\}. For BC and SSM, negative samples are replaced with in-batch samples. The number of layers is 2 for all the LightGCN-based models. We set the max epoch to 1,000 and the early stopping epoch to 10 based on NDCG@20. For their own hyperparameters of each model, we carefully search within the optimum range based on their original papers. For the loss functions with a constant margin, we searched in the range from 0.2 to 1. For our loss, we explored  $\gamma_1$, $\gamma_2$ in the range from 0.1 to 5 for all datasets. All the tables and figures that report only NDCG$@N$ results also show similar trends for Recall$@N$ results. All the results are averaged over 5 runs. The source code is available at \url{https://github.com/psm1206/MAWU}.
\vspace{-1mm}
\renewcommand{\labelenumi}{(\arabic{enumi})}

\section{Experimental results}
\label{sec:result}

This section evaluates the MAWU loss function compared to the baseline loss functions and state-of-the-art CF models. We also validate the effectiveness of margin-aware alignment loss and weighted uniformity loss in MAWU.

\begin{table}[t]
\centering
\caption{Accuracy comparison of ours and CF models on Beauty, Gowalla, and Yelp2018. 
}
\vspace{-4mm}

\label{tab:sota_models} \small
\begin{center}
\renewcommand{\arraystretch}{0.85} 
\newcommand\Tstrut{\rule{0pt}{2.6ex}}         
\begin{tabular}{P{1.2cm}P{2.2cm}P{1.0cm}P{1.0cm}P{1.0cm}}
\toprule
 Dataset & Model & N$@$10 & N$@$20 & N$@$50 \\
\hline
\multirow{10}{*}{Beauty}\Tstrut
 & NeuMF~\cite{HeLZNHC17}  &  0.0373 & 0.0455 & 0.0570  \\
 & ENMF~\cite{ChenZZLM20} &  0.0522 & 0.0621 & 0.0756  \\
 & SimpleX~\cite{MaoZWDDXH21} & 0.0542 & 0.0658 & 0.0813 \\
 \cline{2-5}\Tstrut
 & MultVAE~\cite{LiangKHJ18}  & 0.0438 & 0.0526 & 0.0646  \\
 & RecVAE~\cite{ShenbinATMN20}       & 0.0518 & 0.0615 & 0.0739  \\
\cline{2-5}\Tstrut
 & SGL~\cite{WuWF0CLX21}  & 0.0556 & 0.0670 & 0.0822 \\
 & SimGCL~\cite{YuY00CN22} & 0.0562 & 0.0657 & 0.0773 \\
 & LightGCL~\cite{CaiHXR23} & 0.0575 & 0.0688 & 0.0835 \\
 \cline{2-5}\Tstrut
 & MAWU-MF & 0.0628 & 0.0740 & 0.0881 \\
 & MAWU-LightGCN & \textbf{0.0635} & \textbf{0.0754} & \textbf{0.0905} \\
\hline
\multirow{10}{*}{Gowalla}\Tstrut
 & NeuMF~\cite{HeLZNHC17}   & 0.0884 & 0.1017 & 0.1277  \\ 
 & ENMF~\cite{ChenZZLM20} &  0.1055 & 0.1199 & 0.1475  \\
 & SimpleX~\cite{MaoZWDDXH21} & 0.1179 & 0.1363 & 0.1699  \\
 \cline{2-5}\Tstrut
 & MultVAE~\cite{LiangKHJ18}  & 0.1115 & 0.1281 & 0.1584  \\
 & RecVAE~\cite{ShenbinATMN20}       & 0.1195 & 0.1365 & 0.1672 \\
\cline{2-5}\Tstrut
 & SGL~\cite{WuWF0CLX21}  & 0.1281 & 0.1452 & 0.1772  \\
 & SimGCL~\cite{YuY00CN22} & 0.1226 & 0.1395 & 0.1703 \\
 & LightGCL~\cite{CaiHXR23} & 0.1236 & 0.1417 & 0.1751 \\
 \cline{2-5}\Tstrut
 & MAWU-MF & 0.1277 & 0.1454 & 0.1783 \\
 & MAWU-LightGCN & \textbf{0.1317} &\textbf{0.1496} & \textbf{0.1829} \\
\hline
\multirow{10}{*}{Yelp2018}\Tstrut
 & NeuMF~\cite{HeLZNHC17}  & 0.0417 & 0.0509 & 0.0718 \\
 & ENMF~\cite{ChenZZLM20} &  0.0658	& 0.0768	& 0.1027  \\
 & SimpleX~\cite{MaoZWDDXH21} & 0.0683	& 0.0794	& 0.1047\\
 \cline{2-5}\Tstrut
 & MultVAE~\cite{LiangKHJ18}  & 0.0594	& 0.0714 & 0.0977\\
 & RecVAE~\cite{ShenbinATMN20}       & 0.0689 & 0.0809 & 0.1080 \\
\cline{2-5}\Tstrut
 & SGL~\cite{WuWF0CLX21}  & 0.0668	& 0.0788 & 0.1062 \\
 & SimGCL~\cite{YuY00CN22} & 0.0732	& 0.0848	& 0.1111 \\
 & LightGCL~\cite{CaiHXR23} & 0.0688 & 0.0803 & 0.1074 \\
 \cline{2-5}\Tstrut
 & MAWU-MF & 0.0766 & 0.0894 & 0.1178 \\
 & MAWU-LightGCN & \textbf{0.0779} & \textbf{0.0905} & \textbf{0.1194} \\
\bottomrule
\end{tabular}
\end{center}
\vspace{-3mm}
\end{table}

\begin{table}[t] \small
\centering
\caption{Abalation study on Beauty, Gowalla, and Yelp2018. 
}
\vspace{-4mm}

\label{tab:ablation}
\begin{center}
\renewcommand{\arraystretch}{0.9} 
\newcommand\Tstrut{\rule{0pt}{2.6ex}}         
\begin{tabular}{P{1.2cm}P{1.4cm}P{1.0cm}P{1.0cm}P{1.0cm}}
\toprule
 Dataset & Model & N$@$10 & N$@$20 & N$@$50 \\
\hline
\multirow{3}{*}{Beauty}\Tstrut
 & MAWU & 0.0635 & 0.0754 & 0.0905 \\
 & -MA & 0.0624 & 0.0738 & 0.0883 \\
 & -WU & 0.0628 & 0.0743 & 0.0887 \\
\hline
\multirow{3}{*}{Gowalla}\Tstrut
 & MAWU & 0.1317 & 0.1496 & 0.1829 \\
 & -MA & 0.1315 & 0.1491 & 0.1827 \\
 & -WU & 0.1310 & 0.1482 & 0.1820 \\
\hline
\multirow{3}{*}{Yelp2018}\Tstrut
 & MAWU & 0.0779 & 0.0905 & 0.1194 \\
 & -MA & 0.0778 & 0.0903 & 0.1193 \\
 & -WU & 0.0773 & 0.0898 & 0.1185 \\
\bottomrule
\end{tabular}
\end{center}
\vspace{-3mm}
\end{table}

\begin{table}[t] \small
\centering
\caption{Accuracy comparison with margin types on Beauty. 
}
\vspace{-4mm}

\label{tab:various_margins} 
\begin{center}
\renewcommand{\arraystretch}{0.95} 
\begin{tabular}{P{3.0cm}P{1.1cm}P{1.1cm}P{1.1cm}}
\toprule
 Model & N$@$10 & N$@$20 & N$@$50 \\
\hline
 (1) Zero margin    & 0.0592 & 0.0711 & 0.0872 \\
 (2) Inverse popularity    & 0.0608 & 0.0720 & 0.0865 \\
 (3) UIB fashion~\cite{ZhuoZYZ22}    & 0.0606 & 0.0720 & 0.0867 \\
 (4) BC fashion~\cite{ZhangMWC22}    & 0.0615 & 0.0731 & 0.0871 \\
 (5) Ours    & \textbf{0.0635} & \textbf{0.0754} & \textbf{0.0905} \\
\bottomrule
\end{tabular}
\end{center}
\vspace{-3mm}

\end{table}

\subsection{Performance Comparison}

\noindent
\textbf{Evaluation for baseline loss functions}. Table~\ref{tab:cf_losses} shows experimental results between ten existing loss functions and the MAWU loss function using two backbone models, \ie, MF~\cite{HuKV08} and LightGCN~\cite{0001DWLZ020}, on the three datasets. We report experimental results for each loss category as follows.

\begin{itemize}[leftmargin=5mm]
    \item \textbf{Alignment and uniformity loss functions}: In all the cases, MAWU outperforms DirectAU~\cite{WangYMZCLM22} on average by 3.13\%, 2.11\%, and 1.77\%, respectively. This means that our proposed MAWU is more effective in improving recommendation performance compared to DirectAU. Interestingly, as a dataset becomes sparser (Yelp $\rightarrow$ Gowalla $\rightarrow$ Beauty), the performance gains of MAWU generally increase. This is because MA provides more learning opportunities for unpopular users and items in sparse datasets.

    \vspace{0.5mm}
    \item \textbf{Setwise loss functions}: Except for the results in the Beauty-MF, either CCL~\cite{MaoZWDDXH21} or BC~\cite{ZhangMWC22}, both of which utilize a margin, achieves the best performance. This indicates that incorporating a margin helps improve recommendation performance. Besides, the setwise loss functions (\ie, CCL, SSM, and BC) consistently outperform the pairwise loss functions (\ie, BPR, CML, and SML). This is because the setwise loss functions regard multiple negative items for a single positive item, unlike the pairwise loss functions that only consider a single negative item.
        
   
    \vspace{0.5mm}
    \item \textbf{Pairwise loss functions}: In all cases, BPR~\cite{RendleFGS09} performs better than CML~\cite{HsiehYCLBE17} and SML~\cite{LiZZQZHH20}. This indicates that -LogSigmoid is an effective activation function compared to ReLU. When comparing SML and CML, SML has performance improvements over CML of 29.38\%, 18.87\%, and 18.60\% on average in Beauty, Gowalla, and Yelp2018 for MF backbone, respectively. This suggests that SML's adaptive margin and item-centric loss are used to achieve significant gains.
    
    \vspace{0.5mm}
    \item \textbf{Pointwise loss functions}: We observed two interesting findings about the pointwise loss functions. (i) When LightGCN is the backbone model, UIB is superior to MCL with a constant margin. However, when using MF as the backbone, it shows the opposite trend. This is because the UIB margin is estimated with user embeddings extracted from LightGCN, which is richer than MF. (ii) The pointwise loss functions have on-par performance compared to the setwise loss functions. This is because we set the number of negative samples as high as 30. According to \cite{MaoZWDDXH21}, BCE, one of the pointwise loss functions, performs worse than BPR when the number of negative samples is small while significantly outperforms BPR when the number of negative samples exceeds a certain number (10-20 in the paper).

\end{itemize}

\vspace{1mm}
\noindent
\textbf{Evaluation for CF models}. Table~\ref{tab:sota_models} shows the performance of the proposed model (\ie, MAWU-MF and MAWU-LightGCN) and state-of-the-art MF-, AE-, and GNN-based CF models. Our experimental results show that MF and LightGCN equipped with MAWU consistently outperform other CF models. This indicates that the proposed loss function MAWU leads to better performance, even when used with simple backbone models (\ie, MF and LightGCN). Additionally, we found that GNN-based models (\ie, SGL, SimGCL, and LightGCL) have better performance than MF-based models (\ie, NeuMF, ENMF, and SimpleX) and AE-based models (\ie, MultVAE and RecVAE). This is because GNN-based models utilize high-order relationships from a user-item graph.

\subsection{Breakdown Analysis}
All experiments in this section use LightGCN as a backbone model.

\vspace{1mm}
\noindent
\textbf{Ablation study}. To validate the effectiveness of two components, \ie, margin-aware alignment (MA) and weighted uniformity (WU), we evaluate the performance of ablating MA and WU from MAWU-LightGCN on Beauty, Gowalla, and Yelp2018. Table~\ref{tab:ablation} shows the results of the ablation study. Here, '-MA' and '-WU' are the removal of MA and WU from MAWU, respectively. We can see that eliminating MA and WU leads to a performance drop, indicating that both are effective. Interestingly, on Gowalla and Yelp2018, we found that the performance of MA and WU decreased slightly when only one of them was removed. This suggests that using only one of MA and WU is effective in learning better user and item representations.

\vspace{1mm}
\noindent
\textbf{Comparison with various margin types in MA}. We evaluate the five margin types. The first two types have a constant margin, and the other types learn the margin using UIB~\cite{ZhuoZYZ22}, BC~\cite{ZhangMWC22}, and our fashion (Eq.~\eqref{eq:m_align}), respectively.
 
\begin{enumerate}
    \item Zero margin: $M_{u}$ and $M_{i}=0$, \ie, DirectAU~\cite{WangYMZCLM22}.
    
    \vspace{0.5mm}
    \item Inverse popularity: $M_u = \frac{p_{u,\max}}{p_u}$ and $M_i = \frac{p_{i,\max}}{p_i}$, where $p_u$, $p_i$, $p_{u,\max}$, and $p_{i,\max}$ are user $u$, item $i$ popularity, max popularity of users and items, respectively.

    \vspace{0.5mm}
    \item UIB fashion~\cite{ZhuoZYZ22}: $M_u = W^Tf(u)$ and $M_i = W^Tf(i)$, where $W \in \mathbb{R}^{d \times 1}$ is a learnable parameter.

    \vspace{0.5mm}
    \item BC fashion~\cite{ZhangMWC22}: Its margin is determined as an angle between user and item popularity embedding.
        
    \vspace{0.5mm}
    \item Ours: It learns the margin using Eq.~\eqref{eq:m_align}.
    
\end{enumerate}

Table~\ref{tab:various_margins} shows the highest performance of the type (5) we used. Notably, compared to type (1), types (2) and (4) lead to performance improvement. This indicates that the popularity information helps determine the margin. Type (4) shows better performance than type (2) because it learns the margin. Without explicitly using the popularity information, types (3) and (5) also result in improved performance. Type (3) has a considerable variation according to the backbone model because the margin is obtained by user embedding. However, type (5) learns stably using separate layers and has high performance and fewer parameters than other types. Thus, we adopt type (5) as the margin learning method for the proposed MA.


\begin{figure}
\includegraphics[height=3.4cm]{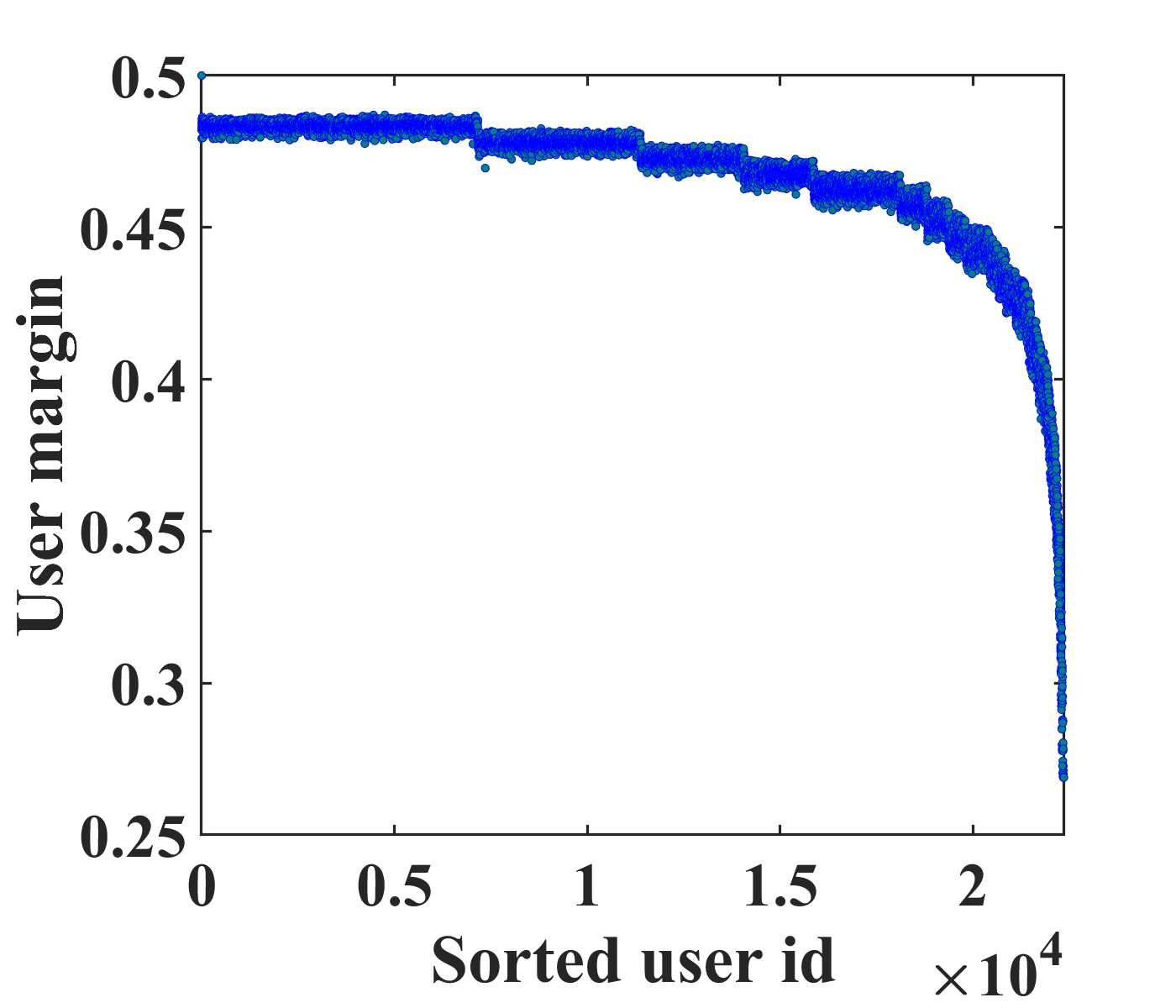} 
\includegraphics[height=3.4cm]{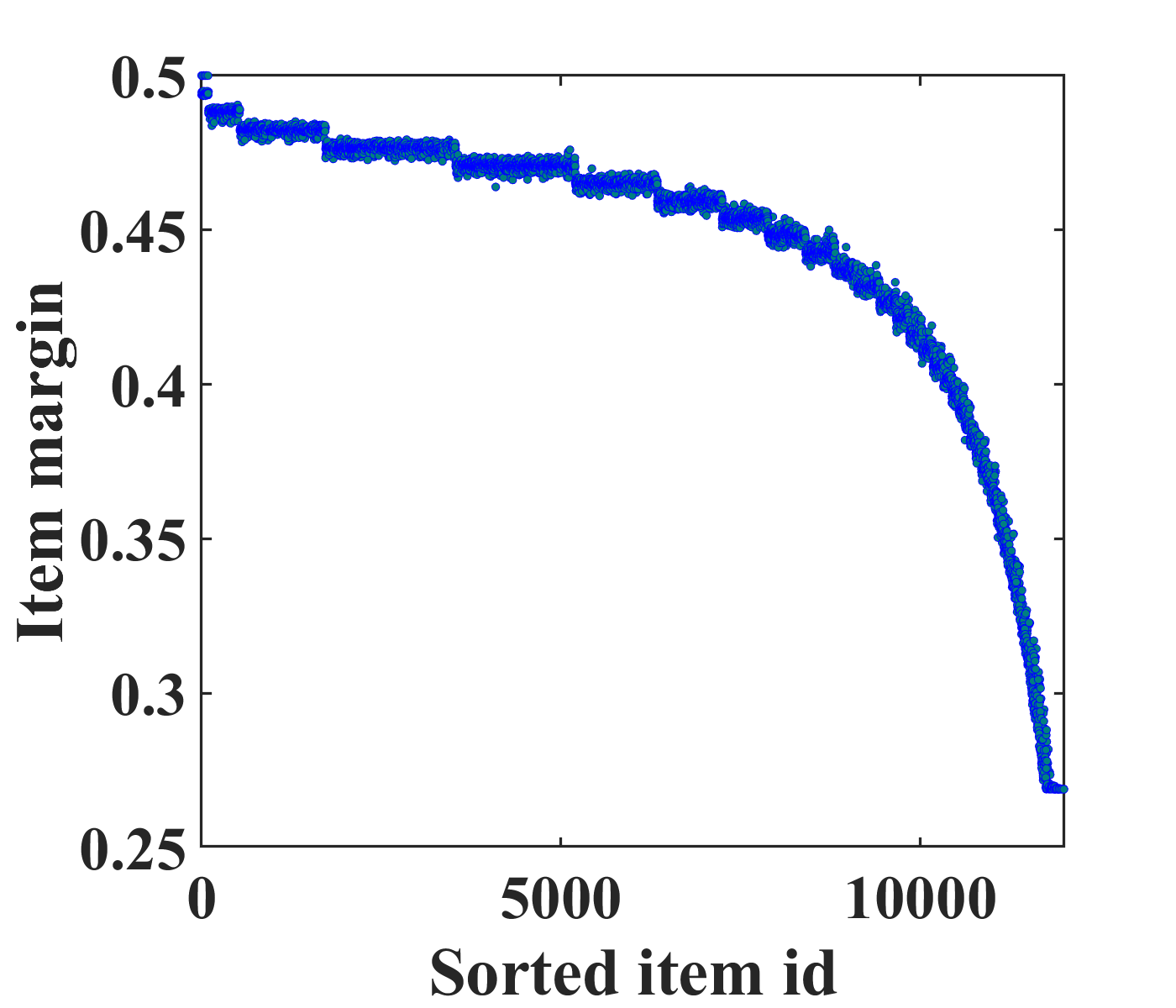} \\ 
\vspace{-2mm}
\caption{Correlation between user/item margin, \ie, $M_u$ and $M_i$, and user/item id sorted by popularity on Beauty. The larger the user/item id, the higher user/item popularity.}\label{fig:user_margin}
\end{figure}



\begin{figure}
\includegraphics[height=2.75cm]{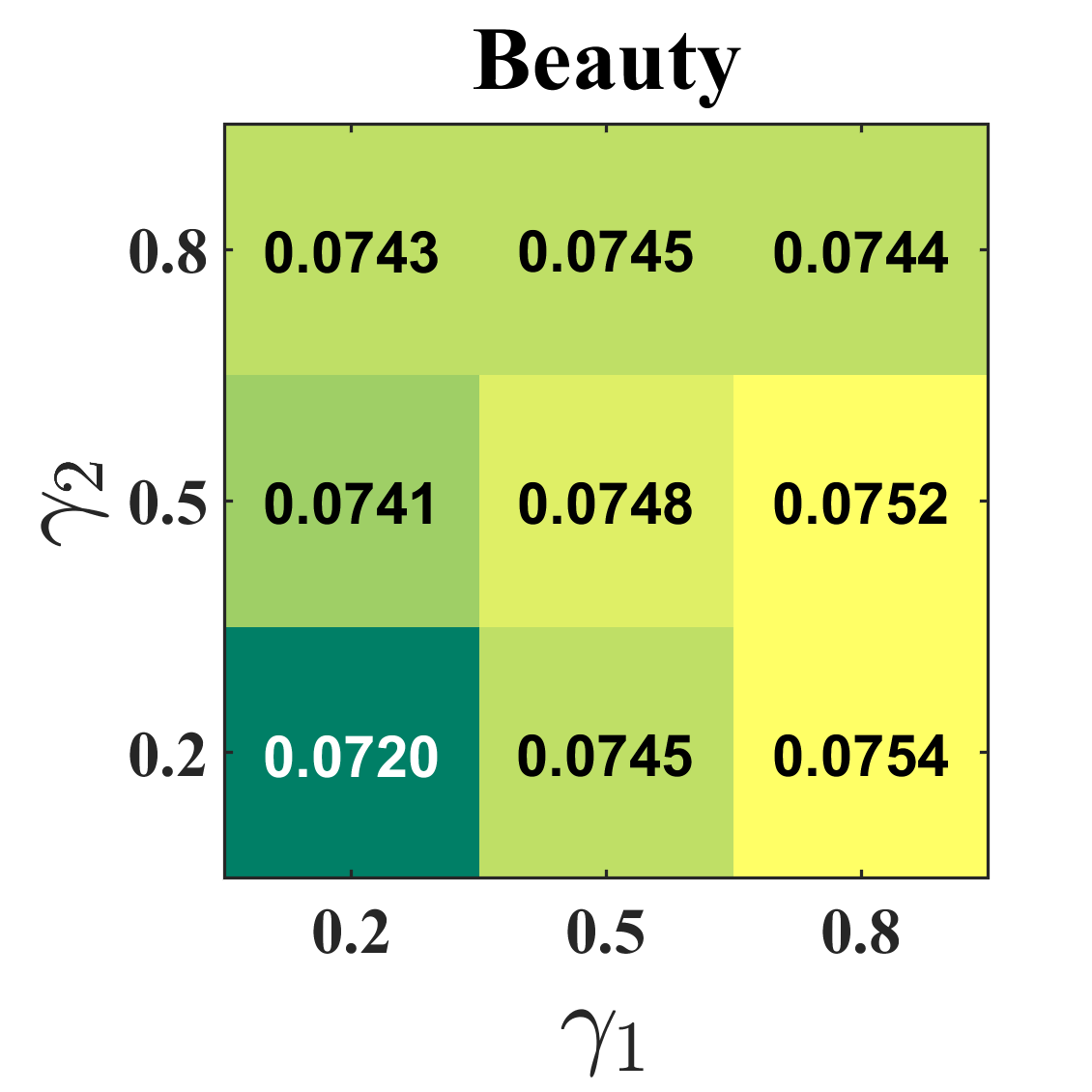} 
\includegraphics[height=2.75cm]{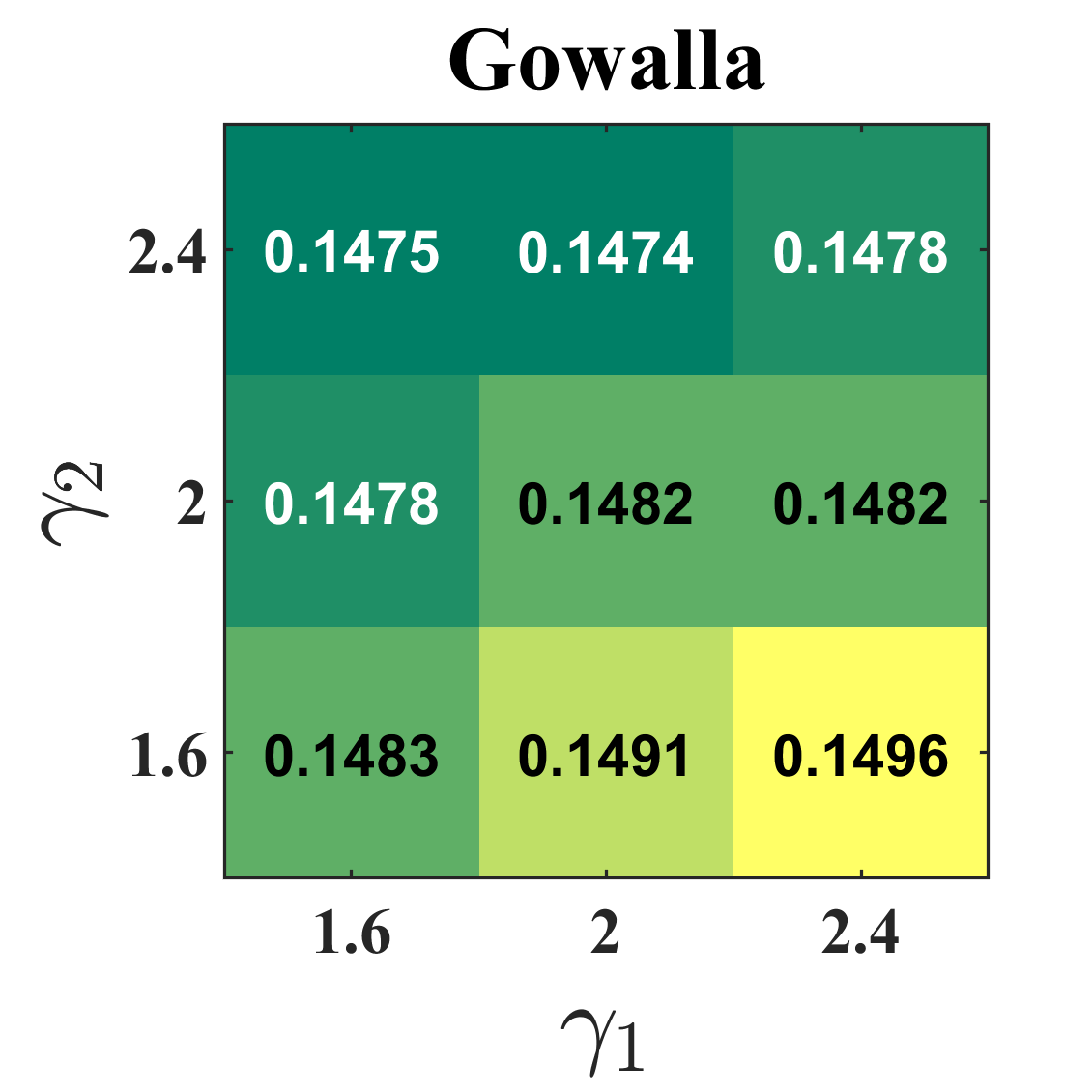} 
\includegraphics[height=2.75cm]{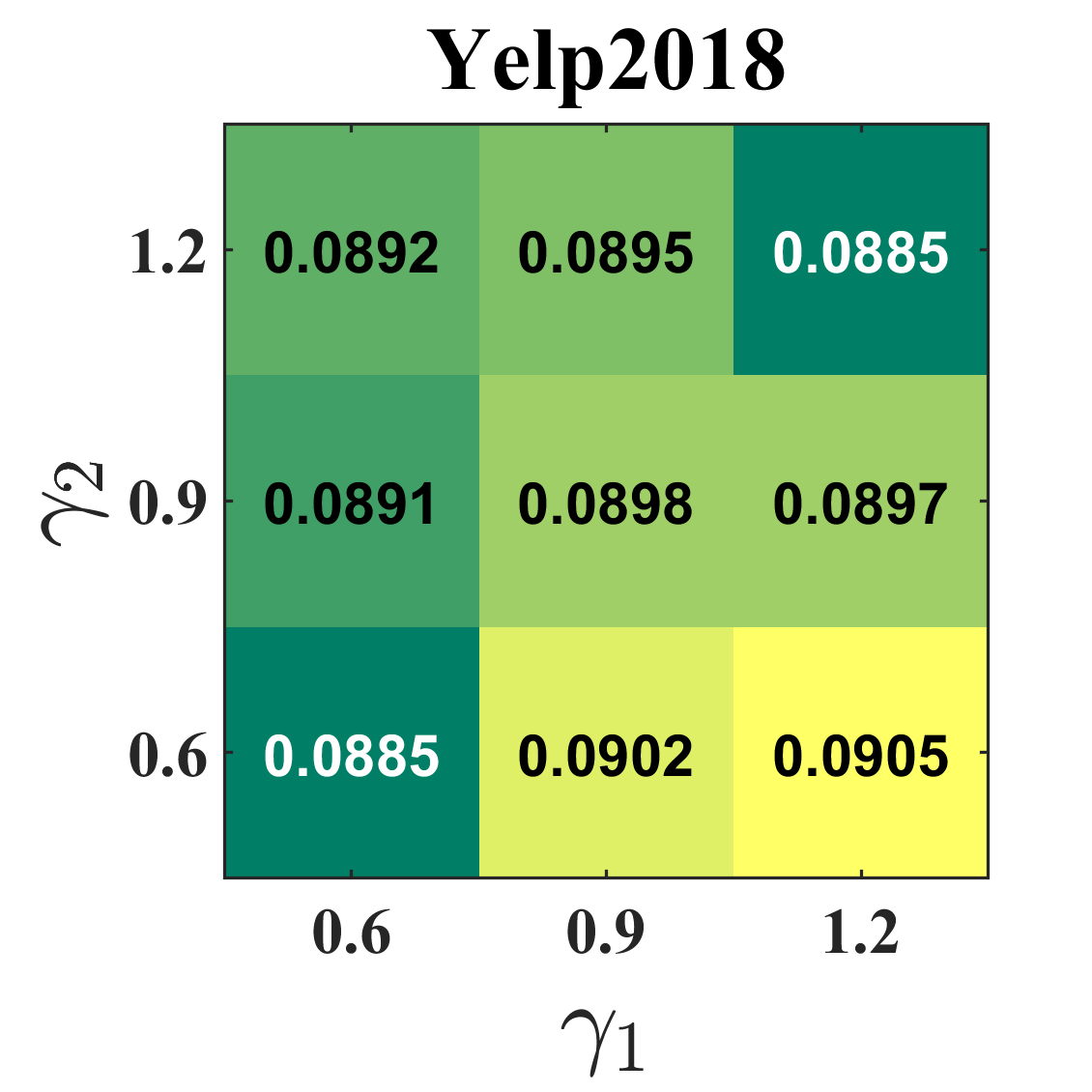} \\ 
\vspace{-2mm}
\caption{Heatmaps of hyperparameters $\gamma_1$ and $\gamma_2$ in Eq.~\eqref{eq:w_unif} on Beauty, Gowalla, and Yelp2018. Each metric is NDCG$@$20.}\label{fig:ui_uniformity}
\end{figure}


\vspace{1mm}
\noindent
\textbf{Margin analysis for user/item popularity}. Figure~\ref{fig:user_margin} depicts that user and item margins learned from MAWU for each user and item sorted by popularity. As we intended, the unpopular users and items have larger margin values. That is, the margin encourages learning for unpopular users and items. Besides, the user and item margins have various margin values within the range of 0.2693-0.4861 and 0.2688-0.4948, respectively. This means that there are suitable margin values for each user and item.

\vspace{1mm}
\noindent
\textbf{Effect of hyperparameters $\gamma_1$ and $\gamma_2$ in WU}.
Figure~\ref{fig:ui_uniformity} presents the results of a grid search for $\gamma_1$ and $\gamma_2$ of the MAWU loss on three different datasets. Lighter colors indicate higher performance in terms of NDCG@20. Our experiments reveal intriguing findings: (i) The performance distribution varies across datasets. In particular, the ratio $\gamma_1 / \gamma_2$ of the best combinations for Beauty, Gowalla, and Yelp2018 are 4, 1.5, and 2, respectively, which is proportional to the ratio of Gini indices in Table~\ref{tab:statistics}. This means that if the item distribution is skewed (large $Gini_i$), the weight for item uniformity should be lowered. (ii) Among all datasets, the highest performance is consistently achieved when $\gamma_1$ is greater than $\gamma_2$. It indicates that achieving uniformity in user representations is more critical than in item representations. From these findings, it is necessary to adjust $\gamma_1$ and $\gamma_2$ properly according to the characteristics of the dataset.

\vspace{-0.5mm}

\section{Related Work}\label{sec:related}

This section reviews existing studies for two parts, \ie, interaction encoder and negative sampling, used in CF models because loss functions are addressed in the previous sections (Refer to Appendix~\ref{appen:related_work} for related work on loss functions). 


\vspace{1mm}
\noindent
\textbf{Interaction encoder.} A lot of existing studies have been actively studied to develop better interaction encoders. It is broadly categorized into two directions: linear and non-linear models. (i) The linear model represents the latent embedding vectors of users and items as a linear combination. It includes neighbor-based regression~\cite{NingK11, Steck19} and matrix factorization~\cite{Koren08, SalakhutdinovM07, ChenZZLM20}. Despite its simplicity, recent studies~\cite{DacremaCJ19, ChinCC22, abs-2305-01801} have shown that linear models often outperform non-linear models. (ii) Non-linear models mostly utilize deep neural networks, such as autoencoders (AE), recurrent neural networks (RNN), graph neural networks (GNN), and so on. Specifically, AE-based models~\cite{SedhainMSX15, LiangKHJ18, ShenbinATMN20} learn complex non-linear relationships among items. RNN-based models~\cite{ManotumruksaMO17, WuABSJ17, HidasiKBT15} adopt the user's sequential behavior information. They are beneficial for sequential and session-based recommender systems. GNN-based models~\cite{0001DWLZ020, WuWF0CLX21, YuY00CN22} capture high-order correlations from a user-item bipartite graph. They learn multi-hop informative relationships, leading to better performances than existing CF models.

\vspace{1mm}
\noindent
\textbf{Negative sampling.} Most CF models utilize implicit feedback, \eg, click and view, which is more applicable to real-world scenarios. So, we need to treat some of the implicit feedback as negative feedback, which is indispensable for pairwise and setwise loss functions. As a sampling method, random sampling~\cite{RendleFGS09} is widely used, but several studies focus on using more informative negative sampling~\cite{WuVSSR19, ChenSSH17, RendleF14, ZhangCWY13, DingQY0J20, YuQ20, DingQ00J19, 0006PZL0Z022, YangYCHLWXC20}. Specifically, \cite{WuVSSR19, ChenSSH17} replaced the uniform distribution with an item popularity-based distribution, and \cite{RendleF14, ZhangCWY13} gave a high sampling probability to negative samples with large prediction scores. Based on the observation that false negative samples have high prediction scores, \cite{DingQY0J20} chooses negative items.

\section{Conclusion}\label{sec:conclusion}

This paper is the first work that mathematically/empirically analyzes existing loss functions. (i) Based on our analysis, we found the potential for model improvement through the integration of BC and DirectAU. (ii) From the potential, we proposed a new loss, called \emph{Margin-aware Alignment and Weighted Uniformity (MAWU)}. In our proposed MAWU, MA encourages CF models to better learn user/item representations by using adaptive margins based on popularity. Then, WU adjusts the importance of user and item uniformity based on dataset statistics, \eg, skewed/uniform distributions. (iii) Extensive experimental results demonstrate that LightGCN equipped with MAWU outperforms existing loss functions and state-of-the-art MF-, AE-, and GNN-based CF models.

\section*{Acknowledgments} 
This work was supported by Institute of Information \& communications Technology Planning \& Evaluation (IITP) grant and National Research Foundation of Korea (NRF) grant funded by the Korea government (MSIT) (No. 2019-0-00421, 2022-0-00680, 2022-0-01045, RS-2023-00219919, and NRF-2018R1A5A1060031).

\newpage
\balance

\newpage

\appendix

\begin{table}[t] \small
\centering
\caption{Accuracy comparison in two user groups (2:8) by popularity. Pop and Unpop mean popular and unpopular user groups, respectively. DAU and MAWU use MF as a backbone. Although we only report NDCG$@N$ results, we observe similar trends with Recall$@N$.
}
\vspace{-2mm}
\label{tab:pop_unpop_user_acc}
\begin{center}
\renewcommand{\arraystretch}{1} 
\begin{tabular}{P{1.2cm}P{1.1cm}P{1.1cm}P{0.9cm}P{0.9cm}P{0.9cm}}
\toprule
 Dataset & Group & Loss & N$@$10 & N$@$20 & N$@$50 \\
\hline
\multirow{6}{*}{Beauty} & \multirow{3}{*}{Pop} 
& DAU &  0.0859 & 0.1047 & 0.1309  \\
& & MAWU  &  0.0952 & 0.1121 & 0.1367  \\
 \cline{3-6}
& & Gain (\%) & 10.83 & 7.07 & 4.43  \\
 \cline{2-6}
& \multirow{3}{*}{Unpop}
& DAU &  0.0534 & 0.0633 & 0.0750  \\
& & MAWU & 0.0555 & 0.0649 & 0.0766  \\
\cline{3-6}
& & Gain (\%) & 3.93 & 2.53 & 2.13 \\
 \cline{1-6}

\multirow{6}{*}{Gowalla} & \multirow{3}{*}{Pop} 
& DAU &  0.1798 & 0.1766 & 0.2086 \\
& & MAWU & 0.1810 & 0.1783 & 0.2099   \\
\cline{3-6}
& & Gain (\%) & 0.67 & 0.96 & 0.62  \\
 \cline{2-6}
& \multirow{3}{*}{Unpop}
& DAU &  0.1124 & 0.1349 & 0.1683  \\
& & MAWU  &  0.1146 & 0.1374 & 0.1710 \\
\cline{3-6}
& & Gain (\%) & 1.96 & 1.85 & 1.60  \\
 \cline{1-6}

\multirow{6}{*}{Yelp2018} & \multirow{3}{*}{Pop} 
& DAU &  0.1284 & 0.1200 & 0.1455  \\
& & MAWU & 0.1317 & 0.1229 & 0.1486  \\
\cline{3-6}
& & Gain (\%) & 2.57 & 2.42 & 2.13  \\
 \cline{2-6}
& \multirow{3}{*}{Unpop}
& DAU &  0.0609 & 0.0786 & 0.1078  \\
& & MAWU  &  0.0623 & 0.0806 & 0.1096  \\
\cline{3-6}
& & Gain (\%) & 2.30 & 2.54 & 1.67  \\
 
\bottomrule
\end{tabular}
\end{center}

\end{table}

\section{Minimization of Margin-aware Alignment Loss Function} \label{appen:proof_ma}
We theoretically prove that the MA loss function that expresses $M_{ui}$ as the sum of $M_u$ and $M_i$ can learn more user/item-specific embeddings than BC~\cite{ZhangMWC22} loss function.

\begin{theorem} \label{minimization_ma}
    Assuming i) the user/item embeddings are normalized, and ii) there are rotation matrices that rotate the user/item embeddings, the minimization of MA loss function is equivalent to minimizing a compactness part. The compactness makes user/item embeddings closer to the averaged item/user embeddings that reflect user/item-specific information (\eg, user/item popularitiess).
\end{theorem}
\begin{align}
\mathcal{L}_{M A} \stackrel{\mathrm{c}}{=} \sum_{u \in \mathcal{U}}\left\|v_u-R_u^{\normalfont \text{T}} c_u^R\right\|^2+\sum_{i \in \mathcal{I}}\left\|v_i-R_i^{\normalfont \text{T}} c_i^R\right\|^2,
\end{align}
where $\stackrel{\mathrm{c}}{=}$ denotes equality up to multiplicative or additive constant. $c_u^R=\frac{1}{\left|\mathcal{P}_u\right|} \sum_{i \in \mathcal{P}_u} R_i v_i$ and $c_i^R=\frac{1}{\left|\mathcal{P}_i\right|} \sum_{u \in \mathcal{P}_i} R_u v_u$ denote average of rotated item and user embeddings, respectively. $\mathcal{P}_u$ and $\mathcal{P}_i$ are a set of items by user $u$ and a set of users by item $i$, respectively. $R_u$ and $R_i$ are rotation matrices of user and item embeddings with the margins $M_u$ and $M_i$, respectively. Besides, $v_u$ and $v_i$ are L2-normalized user and item embedding, respectively~\footnote{For visibility of formulas, we denote $v_u$ and $v_i$ instead of $\Tilde{f(u)}$ and $\Tilde{f(i)}$, respectively.}.

\begin{proof} \label{proof_minimization_ma}
To prove Theorem~\ref{minimization_ma}, we are inspired by a theoretical property of BC~\cite{ZhangMWC22} loss function. The theoretical property means that minimizing BC optimizes both compactness and dispersion simultaneously.

\begin{align}\label{eq:minimization_bc}
    \mathcal{L}_{BC} \geq \sum_{u \in \mathcal{U}}\left\|v_u-c_u\right\|^2+\sum_{i \in \mathcal{I}}\left\|v_i-c_i\right\|^2-\sum_{u \in \mathcal{U}} \sum_{j \in \mathcal{N}_u}\left\|v_u-c_j\right\|^2,
\end{align}
where $c_u=\frac{1}{\left|\mathcal{P}_u\right|} \sum_{i \in \mathcal{P}_u} v_i$ and $c_i=\frac{1}{\left|\mathcal{P}_i\right|} \sum_{u \in \mathcal{P}_i} v_u$ denote averaged item and user embeddings, respectively.

The first two terms are: i) compactness, which makes the user embedding close to the average of the item embeddings, and the last term is: ii) dispersion, which makes the distance between users and items farther apart. Since dispersion implies uniformity and is the corresponding concept of the WU loss function, we aim to verify the compactness of the MA loss function by referring to the first two terms.

Suppose we have rotation matrices (\ie, $R_u$ and $R_i$) that rotate the user/item embeddings by the user/item margins (\ie, $M_u$ and $M_i$), then the MA loss function is derived as follows.
\begin{align}
    \mathcal{L}_{MA}=-\sum_{(u, i) \in \mathcal{D}}\cos \left(\hat{\theta}_{ui}+M_u+M_i\right)=-\sum_{(u, i) \in \mathcal{D}}\left(R_u v_u\right)^{\text{T}}\left(R_i v_i\right).
\end{align}

Here, $v_u$ and $v_i$ are unit vectors (\ie, $v_u^T v_i=1-\frac{1}{2}\left\|v_u-v_i\right\|^2$), which can be expanded as follows.

\begin{align}
& -\sum_{(u, i) \in \mathcal{D}}\left(R_u v_u\right)^{\text{T}}\left(R_i v_i\right) \stackrel{\mathrm{c}}{=} \sum_{(u, i) \in \mathcal{D}}\left\|R_u v_u-R_i v_i\right\|^2 \\
& =\sum_{u \in \mathcal{U}} \sum_{i \in \mathcal{P}_u}\left(\left\|R_u v_u\right\|^2-v_u^{\text{T}} R_u^{\text{T}} R_i v_i\right)+\sum_{i \in \mathcal{I}} \sum_{u \in \mathcal{P}_i}\left(\left\|R_i v_i\right\|^2-v_u^{\text{T}} R_u^{\text{T}} R_i v_i\right) \\
& =\sum_{u \in \mathcal{U}}\left(\left|\mathcal{P}_u\right| v_u^{\text{T}} R_u^{\text{T}} R_u v_u-v_u^{\text{T}} R_u^{\text{T}} \sum_{i \in \mathcal{P}_u} R_i v_i\right) \nonumber \\
& \qquad\qquad\qquad + \sum_{i \in \mathcal{I}}\left(\left|\mathcal{P}_i\right| v_i^{\text{T}} R_i^{\text{T}} R_i v_i-v_i^{\text{T}} R_i^{\text{T}} \sum_{u \in \mathcal{P}_i} R_u v_u\right) \\
& =\sum_{u \in \mathcal{U}}\left(\left|\mathcal{P}_u\right| v_u^{\text{T}}\left(v_u-\frac{R_u^{\text{T}}}{\left|\mathcal{P}_u\right|} \sum_{i \in \mathcal{P}_u} R_i v_i\right)\right) \nonumber \\
& \qquad\qquad\qquad + \sum_{i \in \mathcal{I}}\left(\left|\mathcal{P}_i\right| v_i^{\text{T}}\left(v_i-\frac{R_i^{\text{T}}}{\left|\mathcal{P}_i\right|} \sum_{u \in \mathcal{P}_i} R_u v_u\right)\right) \\
& =\sum_{u \in \mathcal{U}}\left(\left|\mathcal{P}_u\right| v_u^{\text{T}}\left(v_u-R_u^{\text{T}} c_u^R\right)\right) + \sum_{i \in \mathcal{I}}\left(\left|\mathcal{P}_i\right| v_i^{\text{T}}\left(v_i-R_i^{\text{T}} c_i^R\right)\right) \\
& \stackrel{\mathrm{c}}{=} \sum_{u \in \mathcal{U}}\left\|v_u-R_u^{\text{T}} c_u^R\right\|^2+\sum_{i \in \mathcal{I}}\left\|v_i-R_i^{\text{T}} c_i^R\right\|^2. \label{proof_final}
\end{align}

From Eq~\eqref{proof_final}, we can see that the MA loss function makes user embeddings closer to the average of item embeddings reflecting user-specific information (\eg, user popularity) (and vice versa). This suggests that the MA loss function may yield more meaningful embeddings because it can reflect user-specific information compared to simply making user embeddings closer to the average of item embeddings (\ie, BC~\cite{ZhangMWC22}). In fact, Table~\ref{tab:various_margins} shows that the MA loss function method (\ie, type (5)) outperforms the BC method (\ie, type (4)).

\end{proof}

\section{Performance analysis by user popularity}
To analyze where the performance improvement of MAWU comes from, we divide the user groups 2:8 by user popularity to observe the performance. Table~\ref{tab:pop_unpop_user_acc} shows that MAWU outperforms DirectAU~\cite{WangYMZCLM22} in both groups. This seems to depend on the characteristics of the dataset.

\section{Related Work on Various Loss Functions in Collaborative Filtering} \label{appen:related_work}
\textbf{Overview.} Traditional loss functions, \ie,  MSE~\cite{Koren08}, BCE~\cite{HuKV08}, and BPR~\cite{RendleFGS09}, that have been widely used in recommender systems. While many SOTA models still enjoy using these loss functions, a variety of loss functions have been recently proposed in CF~\cite{HsiehYCLBE17, LiZZQZHH20, GaoCPSV22, MaoZWDDXH21, WuWGCFQH22, WangYMZCLM22, ZhuoZYZ22}. These studies are gaining more attention as they show that simply changing the loss function can outperform SOTA models~\cite{MaoZWDDXH21, WuWGCFQH22, GaoCPSV22, WangYMZCLM22}. The research on CF loss function has been developed from the perspective of representation learning, from metric learning~\cite{HsiehYCLBE17, LiZZQZHH20} to contrastive learning~\cite{WuWF0CLX21, YuY00CN22, WuWGCFQH22, ZhangMWC22}. Recently, the decomposition of contrastive loss into alignment and uniformity has shown high performance~\cite{WangYMZCLM22}. In addition, module-type loss function~\cite{GaoCPSV22, ZhuoZYZ22} and representation learning in hyperbolic space~\cite{0001L00K22, 000100LK22} have been proposed. Furthermore, we formally compare the loss functions of graph contrastive learning-based models and traditional models.


\vspace{1mm}
\noindent
\textbf{Metric learning.}
\citet{HsiehYCLBE17} pioneered the development of the loss function, which first introduced the concept of metric learning to CF. CML~\cite{HsiehYCLBE17} incorporates the intuition that the distance between a user and a positive item should be shorter than the distance between a user and a negative item into the loss function. While CML utilizes only user-centric loss, SML~\cite{LiZZQZHH20} considers the concept of item-centric loss additionally. Besides, unlike CML, which uses the margin constant, SML~\cite{LiZZQZHH20} adaptively learns the margin.


\vspace{1mm}
\noindent
\textbf{Contrastive learning.}
The loss functions that are interpreted in terms of contrastive learning have been proposed~\cite{MaoZWDDXH21, WuWF0CLX21, ZhouMZZY21, YuY00CN22, WuWGCFQH22, ZhangMWC22}. A simple form of contrastive loss, CCL~\cite{MaoZWDDXH21}, performs hard negative sampling with a constant margin. In addition, InfoNCE~\cite{abs-1807-03748} and NT-Xent~\cite{ChenK0H20} loss functions, which are popular contrastive loss functions in other domains (\eg, computer vision and natural language processing), are also utilized in recommender systems~\cite{WuWF0CLX21, ZhouMZZY21, YuY00CN22, WuWGCFQH22, ZhangMWC22}. In particular, \citet{WuWGCFQH22} proves that the contrastive loss function mitigates popularity bias and has a hard negative sampling effect, showing its suitability with recommender systems. Motivated by the proof, \citet{ZhangMWC22} presents a new loss function that adds an angular margin to SSM~\cite{WuWGCFQH22}. This margin limits the space where user embeddings can be learned by the angle formed by the popularity embedding of the user and the positive item.


\vspace{1mm}
\noindent
\textbf{Alignment and uniformity.}
Recently, there have been efforts to interpret contrastive loss as two new properties~\cite{WangI20, WangYMZCLM22}. \citet{WangI20} verified that minimizing contrastive loss has the same effect as optimizing alignment and uniformity. Inspired by the proof, \citet{WangYMZCLM22} derives that the BPR loss is smaller than the vanilla BPR loss under perfect alignment and uniformity conditions. Based on the derivation, \citet{WangYMZCLM22} proposed DirectAU~\cite{WangYMZCLM22}, which directly optimizes alignment and uniformity in CF for the first time. DirectAU has the advantage of not requiring negative sampling.


\vspace{1mm}
\noindent
\textbf{Modularization.}
The module-type loss functions~\cite{GaoCPSV22, ZhuoZYZ22} have been proposed that can be added to the existing loss function. Specifically, MCL~\cite{GaoCPSV22} adds several weighting parameters to the original loss functions based on the weight analysis of BPR and triplet loss. \citet{ZhuoZYZ22} introduces a flexible boundary instead of the existing fixed boundary. It considers the flexible boundary as the user interest boundary and divides the positive and negative scores based on that boundary.


\vspace{1mm}
\noindent
\textbf{Hyperbolic space.}
While the previous loss functions all compute in Euclidean space, \citet{000100LK22} first introduces the concept of hyperbolic geometry. The geometric properties of hyperbolic space provide rich information for the model training. HRCF~\cite{000100LK22} computes the distance in hyperbolic space and uses a triplet margin loss. HICF~\cite{0001L00K22} improves the margin learning strategy and negative sampling method in HRCF~\cite{000100LK22} from the perspective of hyperbolic space.


\vspace{1mm}
\noindent
\textbf{Graph contrastive learning}
Lately, Graph Contrastive Learning (GCL)-based models have been demonstrating their effectiveness with high performance~\cite{WuWF0CLX21, YuY00CN22, CaiHXR23}. We thus investigate the mathematical differences between the loss functions of GCL-based models and traditional models. Formally, most GCL-based models use a loss function consisting of two terms: recommendation loss and contrastive loss, \ie, $\mathcal{L}_{GCL} = \mathcal{L}_{Rec} + \lambda \mathcal{L}_{CL}$. The recommendation loss $\mathcal{L}_{Rec}$ is fundamentally used in recommender systems. In other words, $\mathcal{L}_{Rec}$ contains $\mathcal{L}_{Pointwise}$, $\mathcal{L}_{Pairwise}$, $\mathcal{L}_{Setwise}$, and $\mathcal{L}_{AU}$. In \cite{WuWF0CLX21, YuY00CN22, CaiHXR23}, they all use BPR~\cite{RendleFGS09} as $\mathcal{L}_{Rec}$. By adding the contrastive loss $\mathcal{L}_{CL}$ to the existing loss, we can expect to see a performance improvement. The contrastive loss $\mathcal{L}_{CL}$ augments the user and item with two views each and performs contrastive learning on the augmented user and item, \ie, $\mathcal{L}_{CL} = \mathcal{L}_{CL-user} + \mathcal{L}_{CL-item}$. GCL-based models are categorized depending on how the augmentation is designed. SGL~\cite{WuWF0CLX21} uses graph augmentation with node/edge dropout, and SimGCL~\cite{YuY00CN22} utilizes random noise directly on user/item embeddings as an augmentation. Meanwhile, LightGCL~\cite{CaiHXR23} performs SVD on the adjacency matrix and uses the reconstructed graph as an augmentation.


\vspace{1mm}
\noindent
\textbf{Others.}
In addition to designing a single loss function, research on loss function search~\cite{LiYLGWYO19, WangWCZ020} and loss function optimization scheduling~\cite{XuZHLSX19} are also actively studied. Influenced by these studies, recommender systems are also using automatic machine learning to generate loss or customize loss functions~\cite{ZhaoLFLTW21, LiJGZ22, TangBLLZ22}. In line with this trend, this paper aims to be a milestone in the development of CF loss functions.




\end{document}